\global\def\draftcontrol{0}

   \def\versionno{ bulk rn-mass }
\catcode`\@=11

\expandafter\ifx\csname draftcontrol\endcsname\relax\global\def\draftcontrol{0}
\fi

{\count255=\time\divide\count255 by 60
\xdef\hourmin{\number\count255}
\multiply\count255 by-60\advance\count255 by\time
\xdef\hourmin{\hourmin:\ifnum\count255<10 0\fi\the\count255}}
\def\draftdate{\number\month/\number\day/\number\year\ \ \ \hourmin }

\newcommand\makepapertitle{\par
  \begingroup
    \renewcommand\thefootnote{\@fnsymbol\c@footnote}%
    \def\@makefnmark{\rlap{\@textsuperscript{\normalfont\@thefnmark}}}%
    \long\def\@makefntext##1{\parindent 1em\noindent
            \hb@xt@1.8em{%
                \hss\@textsuperscript{\normalfont\@thefnmark}}##1}%
     \newpage
     \global\@topnum\z@   % Prevents figures from going at top of page.
     \@makepapertitle
     \thispagestyle{empty}\@thanks
  \endgroup
  \setcounter{footnote}{0}%
  \global\let\thanks\relax
  \global\let\makepapertitle\relax
  \global\let\@makepapertitle\relax
  \global\let\@thanks\@empty
  \global\let\@author\@empty
  \global\let\@date\@empty
  \global\let\@title\@empty
  \global\let\title\relax
  \global\let\author\relax
  \global\let\date\relax
  \global\let\and\relax
  \def\version{\let\version\@version\@gobble}
}
\def\@makepapertitle{%
  \newpage
   \ifnum\draftcontrol=1 {}
   \version\versionno
   \vskip 3em%
   \else
   \hfill\hbox to 3cm {\parbox{4cm}{\@pubnum}\hss}%
   \vskip 3em%
   \fi
   \begin{center}%
   \let \footnote \thanks
     {\LARGE {\@title}}%
     \vskip 1.5em%
     {\normalsize%\large
       \lineskip .5em%
       \begin{tabular}[t]{c}%
         \@author
       \end{tabular}\par}%
     \vskip 1.5em%
     {\@bstract}%
     \end{center}%
     \vskip 1.5em
     \@date%
   \par
}

\gdef\@pubnum{}
\def\pubnum#1{%
  \gdef\@pubnum{#1}}

\gdef\@bstract{}
\def\Abstract#1{%
  \gdef\@bstract{%
   \parbox{\textwidth-0pc}{%
   \centerline{\bf Abstract}\penalty1000%
\kern.2cm%
\noindent%\abstractfont \baselineskip=12pt
\renewcommand\baselinestretch{1.0}%
{#1}}}
}

\def\ps@paper{\let\@mkboth\@gobbletwo%
     \ifnum\draftcontrol=1
    \def\@oddfoot{\hbox to \textwidth{\tiny \versionno \hfil\tiny\draftdate}%
    \hskip -\textwidth \hbox to \textwidth{\hfil\rm\thepage\hfil}}%
     \else\def\@oddfoot{\hbox to \textwidth{\hfil\rm\thepage\hfil}}
     \fi
     \let\@evenfoot\@oddfoot
}
\def\body{\clearpage
          \pagestyle{paper}
    }
\def\@version#1{\ifnum\draftcontrol=1
\typeout{}\typeout{#1}\typeout{}
\vskip3mm\centerline{\hbox{\fbox{\normalsize{\tt DRAFT -- #1 -- }
                   {\draftdate}}}}\vskip3mm
\fi}
\let\version\@version
\long\def\eqlabel#1{\ifnum\draftcontrol=1
                    \tag@false  % there are some problems with multline without this
                    \tag*{(\theequation) \hbox to -0.2cm{\hspace{0cm}\small{#1}\hss}}
                    \refstepcounter{equation}
                    \edef\@currentlabel{\theequation}
                    \ltx@label{#1}          % use old LaTeX \label instead of new definition
                    \else
                    \label{#1}
                    \fi
                    }
\let\st@bibitem\@bibitem
\let\st@lbibitem\@lbibitem
\ifnum\draftcontrol=1
  \def\@bibitem#1{%
    \st@bibitem{#1}\a@@label{#1}\ignorespaces}
  \def\@lbibitem[#1]#2{%
    \st@lbibitem[#1]{#2}\a@@label{#2}\ignorespaces}
  \def\a@@label#1{%
    \gdef\a@lab{\smash{\normalfont\small#1}}
    \ifvmode
      \if@inlabel
        \global\setbox\@labels\hbox{%
          \llap{\a@lab\let\a@lab\relax
                \kern\@totalleftmargin\kern\marginparsep}%
          \box\@labels}%
      \fi
    \fi}
\fi
\documentclass[12pt,letterpaper]{article}
\usepackage{amsmath,amssymb,array,calc,epsfig}
\usepackage{graphicx}
\usepackage{psfrag,verbatim,bm}
\ifnum\draftcontrol=1
\fi
\renewcommand\baselinestretch{1.25}
\renewcommand\section{\@startsection {section}{1}{\z@}%
                                   {-3.5ex \@plus -1ex \@minus -.2ex}%
                                   {2.3ex \@plus.2ex}%
                                   {\normalfont\large\bfseries}}
\renewcommand\subsection{\@startsection{subsection}{2}{\z@}%
                                   {-3.25ex\@plus -1ex \@minus -.2ex}%
                                   {1.5ex \@plus .2ex}%
                                   {\normalfont\normalsize\bfseries}}
\renewcommand\subsubsection{\@startsection{subsubsection}{3}{\z@}%
                                   {-3.25ex\@plus -1ex \@minus -.2ex}%
                                   {1.5ex \@plus .2ex}%
                                   {\normalfont\normalsize\it}}
\renewcommand\paragraph{\@startsection{paragraph}{4}{\z@}%
                                   {-3.25ex\@plus -1ex \@minus -.2ex}%
                                   {1.5ex \@plus .2ex}%
                                   {\normalfont\normalsize\bf}}

\numberwithin{equation}{section}

\def\ie{{\it i.e.}}

\def\revise#1       {\raisebox{-0em}{\rule{3pt}{1em}}%
                     \marginpar{\raisebox{.5em}{\vrule width3pt\
                     \vrule width0pt height 0pt depth0.5em
                     \hbox to 0cm{\hspace{0cm}{%
                     \parbox[t]{4em}{\raggedright\footnotesize{#1}}}\hss}}}}

\newcommand\nxt[1]  {\\\fnxt#1}

\def\cala         {{\cal A}}

\def\calc         {{\cal C}}
\def\cald         {{\cal D}}
\def\cale         {{\cal E}}

\def\cali         {{\cal I}}
\def\calj         {{\cal J}}

\def\call         {{\cal L}}
\def\calm         {{\cal M}}
\def\caln         {{\cal N}}
\def\calo         {{\cal O}}

\def\calr         {{\cal R}}

\def\calz         {{\cal Z}}

\def\del          {\partial}

\def\tr           {\mathop{\rm Tr}}

\def\Im           {{\rm Im\hskip0.1em}}

\def\sqr#1#2{{\vcenter{\vbox{\hrule height.#2pt
 \hbox{\vrule width.#2pt height#1pt \kern#1pt
 \vrule width.#2pt}\hrule height.#2pt}}}}

\def\a{\alpha}
\def\b{\beta}
\def\l{\lambda}
\def\w{\omega}
\def\dd{\delta}
\def\r{\rho}
\def\c{\chi}
\newcommand{\qq}{\mathfrak{q}}
\newcommand{\ww}{\mathfrak{w}}

\def\k{\kappa}
\def\hr{\hat{r}}
\def\om{\Omega}
\def\e{\epsilon}
\def\bm{\bar{\mu}}
\def\tc{\tilde{\calc}}
\def\tz{\tilde{z}}
\def\hz{\hat{z}}
\def\t{\tau}

\catcode`\@=12

\begin{document}

%%%
%%%%%% text starts here
%%%%%%%%%

\title{\bf Critical phenomena in $\caln=4$ SYM plasma}
\pubnum
{UWO-TH-10/01
}

\date{May 2010}

\author{
Alex Buchel\\[0.4cm]
\it Department of Applied Mathematics\\
\it University of Western Ontario\\
\it London, Ontario N6A 5B7, Canada\\
\\
\it Perimeter Institute for Theoretical Physics\\
\it Waterloo, Ontario N2J 2W9, Canada
\\
\\
Albert Einstein Minerva Center\\
 Weizmann Institute of Science\\
Rehovot 76100, Israel\\
}

\Abstract{
Strongly coupled $\caln=4$ supersymmetric Yang-Mills plasma at finite
temperature and chemical potential for an R-symmetry charge undergoes
a second order phase transition. We demonstrate that this phase
transition is of the mean field theory type. We explicitly show that
the model is in the dynamical universality class of 'model B'
according to the classification of Hohenberg and Halperine, with
dynamical critical exponent $z=4$.  We study bulk viscosity in the
mass deformed version of this theory in the vicinity of the phase
transition.  We point out that all available models of bulk viscosity
at continuous phase transition are in conflict with our explicit
holographic computations.
}

\makepapertitle

\body

\version\versionno

\section{Introduction}
According to gauge theory/string theory correspondence of
Maldacena \cite{m9711} maximally supersymmetric $\caln=4$ $SU(N)$
Yang-Mills (SYM) theory is dual to string theory on $AdS_5\times S^5$.
In the planar limit ($g_{YM}^2\to 0$, $N\to\infty$ with $\l\equiv
g_{YM}^2 N$ kept fixed) and for large t' Hooft coupling $\l\gg 1$ the
strongly coupled SYM is described by classical type IIB supergravity
on $AdS_5\times S^5$, making it essentially soluble. The value of this
holographic duality is that it can provide explicit tests of various
phenomenological models invented to describe the dynamics of strongly
coupled systems. The focus of this paper is the application of
gauge/gravity duality to the transport properties of strongly coupled
gauge theory plasma in the vicinity of the second order phase
transitions.

In \cite{bp3}  it was argued that the only model for the critical behavior of the 
bulk viscosity in strongly coupled systems at continuous phase transitions not in conflict 
with explicit holographic computations was that proposed by Onuki \cite{bulk3}.
Specifically, Onuki's model predicts that close to the phase transition the bulk viscosity 
scales as 
\begin{equation}
\zeta\propto |t|^{-z\nu+\a}\,,
\eqlabel{oscale}
\end{equation}
where 
\begin{equation}
t\equiv \frac{T}{T_c}-1 \,,
\eqlabel{redt}
\end{equation}
is the reduced temperature, $\nu$ and $\a$ are the usual static critical exponents of the continuous
 phase transition, and $z$ is a dynamical critical exponent. In this paper we would like to definitely 
answer the question as to whether or not \eqref{oscale}  is realized 
in a strongly coupled gauge theory plasma with 
a holographic dual.

Our starting point is the best studied example of gauge theory/string theory duality, namely 
that of $\caln=4$ SYM plasma. This theory has an $SO(6)\sim SU(4)$ R-symmetry; thus one can turn 
on three independent chemical potentials  (one for each of the $U(1)$'s in 
the Cartan subalgebra of the R-symmetry group).
It is well known that $\caln=4$ SYM plasma at finite temperature $T$ and for a single 
$U(1)$ R-symmetry\footnote{This is not the diagonal $U(1)$ of the $SU(4)$ R-symmetry.} 
chemical potential $\mu$ undergoes a second order phase transition\footnote{As we show below, 
some of the static critical 
exponent first computed in \cite{stu2} and since then widely used in the literature 
are incorrect. This issue could be traced back to the fact that the hyperscaling relation 
between static critical exponents is violated in this theory.} \cite{stu1,stu2,stu3}.   
Moreover, recently \cite{maeda1}, the conductivity $\sigma_Q$ of this gauge theory plasma was shown to be finite 
on the critical line 
\begin{equation}
\frac{\mu}{T}\bigg|_{critical}=\frac{\pi}{\sqrt{2}}\,.
\eqlabel{tmucritical}
\end{equation} 
As a result, the authors of \cite{maeda1} argued that the dynamical universality class of 
$\caln=4$ SYM plasma is that of 'model B' according to classification of Hohenberg and Halperine
\cite{hh}, with the dynamical critical exponent 
\begin{equation}
z=4-\eta\,,
\eqlabel{zma}
\end{equation} 
with $\eta$ being the anomalous static critical exponent.
 In this paper we confirm 
the identification made in \cite{maeda1}, and compute $z$ for the $\caln=4$ SYM plasma. 

Unfortunately, we can not use the $\caln=4$ SYM plasma directly to test Onuki's prediction 
for the scaling of the bulk viscosity in the vicinity of the phase 
transition \eqref{oscale} --- conformal invariance of the theory guarantees that the bulk 
viscosity must vanish for arbitrary chemical potential and the temperature.
Thus, we need to deform the theory in such a way that we break the scale invariance. 
The simplest deformation one can consider is to give mass $M$ to fermions of $\caln=4$ SYM. 
If 
\begin{equation}
{M}\ \ll\ {T_{critical}}\,,
\eqlabel{massfer}
\end{equation}
it is sufficient to work to  order  $\calo\left(\frac{M^2}{T^2}\right)$.
Although not necessary, one can think about above deformation (to the order specified) 
as that  corresponding to deforming  $\caln=4$ plasma to $\caln=2^*$ plasma \cite{pw1,pw2,pw3}.
 
The paper is organized as follows. We being section 2 with presenting the 
effective action for the holographic dual of $\caln=4$ SYM plasma, deformed 
by a dimension-3 operator. We study this gravitational model at finite temperature and  
chemical potential to order $\calo\left(\frac{M^2}{T^2}\right)$: we (numerically) determine the background geometry,
discuss the holographic renormalization of the theory, and compute the equilibrium 
thermodynamics. We present a highly nontrivial test on our analysis by demonstrating 
that the basic thermodynamic relations are satisfied.
Appendix A contains necessary technical details.
 In section 3 we study 
hydrodynamic fluctuations in charged gauge theory plasma and their holographic dual ---
the lowest quasinormal modes of the (mass-deformed) Reissner-Nordstr\"om (RN)
asymptotically $AdS_5$ black holes. We derive the speed of sound and the 
sound wave attenuation coefficient in a generic charged plasma. 
We find that the attenuation coefficient is sensitive to both the shear and 
the bulk viscosities of the plasma, as well as the plasma conductivity. 
Interestingly, the dependence on a conductivity in the attenuation coefficient 
arises only at the fourth order in the parameter breaking the scale invariance, 
\ie, $\frac{M}{T}$. We derive (coupled) equations for the quasinormal modes in the deformed 
RN black hole background and explain how to decode from their spectrum the speed of sound 
waves and the bulk viscosity of the dual plasma. We point out the computational difficulty in
the numerical analysis, intrinsic to finding  the sound channel quasinormal modes in charged 
black hole backgrounds, and present a new method of computing the corresponding 
quasinormal modes. Appendix B is used to explain this new method in a simple setting 
of $\caln=4$ SYM plasma at finite temperature, but zero chemical potentials.
As a highly nontrivial test of our analysis, we show that the speed of sound 
obtained from the thermodynamic analysis is in excellent agreement with the 
one extracted from the dispersion relation of the lowest sound  channel
quasinormal mode. In section 4 we discuss dynamical critical phenomena in 
$\caln=4$ SYM plasma. We review how the dynamical susceptibility can be used 
to compute static anomalous critical exponent $\eta$ and the critical exponent $\nu$ associated 
with the divergence of the correlation length in the vicinity of the transition,
as well as the dynamical critical exponent $z$. We 
explain how the problem of finding the sound channel quasinormal mode 
can be adjusted to extract the dynamical susceptibility. We emphasize 
why the latter analysis can not be performed in the hydrodynamic limit. 
We present results for the critical exponents for $\caln=4$ SYM plasma 
and  demonstrate that these exponents are robust against the mass deformation of 
$\caln=4$ plasma.  We summarize and interpret all the results in concluding section 5.

\section{Holographic dual of mass deformed $\caln=4$ plasma at equilibrium}
Effective five-dimensional action 
describing the holographic dual to $\caln=4$ SYM deformed by an operator $\calo_\Delta$ of dimension $\Delta$
takes form\footnote{The supergravity gauge coupling is chosen so that the asymptotic $AdS_5$ radius 
is one.}
\begin{equation}
\begin{split}
S_5=&\frac{1}{16\pi G_5} \int_{\calm_5}d^5\xi \sqrt{-g}\ \call\\
=&\frac{1}{16\pi G_5} \int_{\calm_5}d^5\xi \sqrt{-g}\left(R-\frac 14 \phi^{4/3} F^2 -
\frac{1}{3}\phi^{-2}\left(\del\phi\right)^2+4\phi^{2/3}+8\phi^{-1/3}+\dd\call\right)\,,
\end{split}
\eqlabel{ac5}
\end{equation}
where $\dd\call$ is a mass deformation
\begin{equation}
\dd\call=-\frac 12 \left(\del\chi\right)^2-\frac{m^2}{2}\chi^2+\calo\left(\c^4\right)\,.
\eqlabel{dldef}
\end{equation}
As usual, the mass of  $\chi$ is related to the dimension $\Delta$ of the corresponding operator in the dual
description
\begin{equation}
\Delta (\Delta-4)=m^2\,.
\eqlabel{md}
\end{equation} 
In what follows we focus on $\Delta=3$ ($m^2=-3$) deformation.

The non-normalizable component $\lambda$ of $\chi$ near the (asymptotic) 
$AdS_5$ boundary is related to the coupling $M$ of operator $\calo_3$ 
deforming the Lagrangian $\call_{CFT}$ of the $\caln=4$ conformal fixed point:
\begin{equation}
\call_{CFT}\to \call_{CFT}-M \calo_3\,,\qquad M\propto \lambda\,,
\eqlabel{cftdef}
\end{equation}  
where the precise definition of $\lambda$ and the relation between 
$M$ and $\lambda$ (up to an irrelevant $c$-number 
normalization of $\calo_3$) will be established later. Notice that 
the identification of \eqref{cftdef} with the dual holographic
action \eqref{ac5}  can be established only to 
order $\l^2$, and thus $\chi^2$ in  \eqref{dldef}. The latter fact is emphasized 
by an $\calo(\c^4)$ term in \eqref{dldef}. The reason why this is so 
is best illustrated with the $\caln=2^*$ example of the holographic gauge 
theory/string theory correspondence \cite{n24}. The duality studied in \cite{n24}
is one of the few examples where it is possible to match exactly the gravitational 
parameters with the corresponding gauge-theoretical ones. In this case, 
a simple deformation of the type \eqref{cftdef} leads to a complicated 
potential for a supergravity scalar $\chi$ --- nonetheless, up to 
order $M^2$, the dual supergravity deformation is unambiguously fixed 
by the scaling dimension of $\calo_\Delta$, as in \eqref{dldef}. 

We are interested in the critical phenomena in $\caln=4$ plasma (and its massive deformation) 
at finite temperature and chemical potential --- thus as far as $M\ll T$, in particular for $T=T_c$, 
if is sufficient to work with the effective holographic description \eqref{ac5}, \eqref{dldef}.

\subsection{Background}
In this section we set up our notations for describing the background geometry of 
$\caln=4$ SYM plasma at finite temperature and  
$U(1)_R$ chemical potential, deformed by $\calo_3$ operator to quadratic order in its coupling.

Consider the following ansatz
\begin{equation}
ds_5^2=-c_1^2\ dt^2+c_2^2\ d\vec{x}^2+c_3^2\ dr^2\,,\qquad A_\mu=A\ \dd_\mu^t\,,
\eqlabel{back}
\end{equation}
where $c_i=c_i(r)$, $A=A(r)$, $\phi=\phi(r)$, $\chi=\chi(r)$.
We find it convenient to introduce a new radial coordinate $x$ as follows
\begin{equation}
1-x=\frac{c_1}{c_2}\,,
\eqlabel{defx}
\end{equation}
so that $x\to 0_+$ corresponds to an AdS boundary and $x\to 1_-$ corresponds to a regular 
Schwarzschild horizon.  Further, we introduce 
\begin{equation}
\phi= H^3\,,\qquad c_2=g\ \left(\frac{H}{H^3-1}\right)^{1/2}\,,
\eqlabel{c2def}
\end{equation}       
with $g=g(x)$, $H=H(x)$.
We would like to construct background geometry perturbatively in $\chi$, 
in other words we parametrize the background as 
\begin{equation}
\begin{split}
g=g_0+\l^2\ g_2\,,\qquad H=H_0+\l^2\ H_2\,,\qquad A=A_0+\l^2\ A_2\,,\qquad \chi=\l \chi_1\,,
\end{split}
\eqlabel{defback}
\end{equation} 
where the coefficient of the non-normalizable mode of $\chi$, \ie, $\lambda$,  is introduced so that 
\begin{equation}
\chi_1=x^{1/4}+\calo(x^{1/2})\,,\qquad {\rm as}\qquad x\to 0_+\,.
\eqlabel{deflambda}
\end{equation}
\nxt To order $\calo(\l^0)$ we find
\begin{equation}
\begin{split}
&g_0=\b\,,\ \qquad A_0=\frac{\b\sqrt{1+\k}}{\k}\ \left(\frac{1}{H_0^{3}}-\frac{1}{1+\k}\right)\,,\\
&H_0=\frac{2(1+\k)+(2x-x^2)\k^2+\k\sqrt{x(2-x)(2+x\k)(2(1+\k)-x\k)}}{2(1+\k)}\,,
\end{split}
\eqlabel{zeroback}
\end{equation}  
where the two constants $\{\b,\k\}$ are related to the temperature $T$ and the chemical potential $\mu$ of the 
R-charged black brane
\begin{equation}
2\pi T\bigg|_{\l=0}=\b\ \frac{\k+2}{\sqrt{\k(1+\k)}}\,,\qquad \mu\bigg|_{\l=0}= \frac{\b}{\sqrt{{1+\k}}}\,.
\eqlabel{tmu0}
\end{equation}
Note that the ratio 
\begin{equation}
\frac{2\pi T}{\mu}\bigg|_{\l=0}=\sqrt{{\k}}+\frac{2}{\sqrt{{\k}}}
\eqlabel{ratio0}
\end{equation}
attains a minimum at $\k(\l=0)=2$, at which the black brane undergoes a second order phase transition \cite{stu1,stu2,stu3}.
\nxt To order $\calo(\l^1)$ we have
\begin{equation}
\begin{split}
0=&\c_1''+\frac{H_0^9 (1+\k)+H_0^6 (1+\k)+H_0^3 (2 \k^2-\k+2 \k^2 x^2-1-4 x \k^2)-1-\k}{(H_0^6-1) (H_0^3+1) (x-1) (1+\k)}
\c_1'\\
&-\frac{H_0^7 \k^2 m^2 (H_0^3 (\k^2 x^2-2 x \k^2+H_0^3-2 \k+\k^2-2)+\k+1+H_0^6 \k)}{(1+\k)^2 (H_0^6-1)^2 (H_0^3-1)^2}\c_1\,.
\end{split}
\eqlabel{eq4}
\end{equation}
\nxt To order $\calo(\l^2)$ we have
\begin{equation}
\begin{split}
0=&g_2''+\calc_{11}\ g_2'+\calc_{12}\ H_2'+\calc_{13}\ A_2'+\calc_{14}\ (\c_1')^2+\calc_{15}\ g_2+\calc_{16}\ H_2+\calc_{17}\ \chi_1^2\,,\\
0=&H_2''+\calc_{21}\ g_2'+\calc_{22}\ H_2'+\calc_{23}\ A_2'+\calc_{24}\ (\c_1')^2+\calc_{25}\ g_2+\calc_{26}\ H_2+\calc_{27}\ \chi_1^2\,,\\
0=&A_2''+\calc_{31}\ g_2'+\calc_{32}\ H_2'+\calc_{33}\ A_2'+\calc_{34}\ g_2+\calc_{35}\ H_2\,,
\end{split}
\eqlabel{eq3}
\end{equation}
where the coefficients $\calc_{ij}$ are collected in Appendix \ref{appa}.

\subsection{Holographic renormalization and the boundary stress-energy tensor}

In this section we carefully perform the holographic renormalization of \eqref{ac5} and extract the thermodynamic quantities, subject to 
the following constraints:
\nxt the background geometry of the gauge theory dual to the gravitational action \eqref{ac5} is $R^{3,1}$;
\nxt since \eqref{ac5} is expected to be valid only to order $\calo(\chi^2)$, we perform holographic renormalization 
only to this order in $\chi$, but exact in $\phi$.

Let $r$ be the position of the boundary, and $S_E^r$ be the Euclidean gravitational action on the cut-off space 
\begin{equation}
\lim_{r\to \infty} S_E^r=S_E\,,
\eqlabel{ser}
\end{equation}
where $S_E$ is the Euclidean version of \eqref{ac5}. Explicitly, using equations of motion,
\begin{equation}
\begin{split}
S_E^r=&\frac{1}{16\pi G_5}\int_{r_h}^r dr\int_{\del\calm}d^4\xi \sqrt{h_E}\call_E
=-\frac{1}{16\pi G_5}\int_{r_h}^r dr\int_{\del\calm}d^4\xi \sqrt{h}\call\\
=&\frac{1}{8\pi G_5}\ \left(\int_{\del\calm_5}d^4\xi\right)\  \int_{r_h}^{r}dr \left[\frac{c_1c_2^2c_2'}{c_3}\right]'
=\frac{1}{8\pi G_5}\ \left(\int_{\del\calm_5}d^4\xi\right)\ \times \left[\frac{c_1c_2^2c_2'}{c_3}\right]\bigg|^r_{r_h}\,,
\end{split}
\eqlabel{s5bulk}
\end{equation}
where $r_h$ is a position of the regular Schwarzschild horizon of \eqref{back}, $h_E$ is the induced 
metric on the boundary,  and 
\begin{equation}
\frac{1}{8\pi G_5}=\frac{N^2}{4\pi^2}\,.
\eqlabel{g5}
\end{equation} 
Notice that for a regular Schwarzschild horizon, the horizon contribution in \eqref{s5bulk} vanishes. 
Besides the standard Gibbons-Hawking term 
\begin{equation}
S_{GH}=-\frac{1}{8\pi G_5}\ \int_{\del\calm_5}d^4\xi \sqrt{h_E}\nabla_\mu n^\mu=-\frac{1}{8\pi G_5}
\frac{\left[c_1 c_2^3\right]'}{c_3}\int_{\del\calm_5}d^4\xi\,,
\eqlabel{gh}
\end{equation}
we supplement the combined regularized action $(S_E^r+S_{GH})$ by the appropriate boundary 
counterterms which are needed to get the finite action. These boundary counterterms 
must be constructed from the local metric and $\{\phi,\chi\}$ scalar invariants\footnote{In principle,
there could be a (finite) contribution from the gauge field $A_\mu$.  Such contribution was shown in 
\cite{ls} to vanish.} 
\begin{equation}
S^{counter}=\frac{1}{4\pi G_5}\int_{\del\calm_5}d^4\xi \sqrt{h_E}
\biggl(\a_1+\a_2\ \chi^2+\a_3\ \left(\ln \phi\right)^2+\calo(\chi^4)\biggr)\,,
\eqlabel{scounter}
\end{equation}
$\a_i$ are constant coefficients of the counterterms which are determined by the requirement of having 
a finite action. The counterterm $\propto \a_3$ is finite, but is required to insure the 
first law of thermodynamics \cite{bp} and the supersymmetry (at extremality) \cite{ls} of the
single-charge $AdS_5$ RN black hole.  Thus, the finiteness of the renormalized Euclidean action  
\begin{equation}
\cali_E\equiv \lim_{r\to \infty}\biggl(S_E^r+S_{GH}+S^{counter}\biggr)\,,\qquad |\cali_E|<\infty\,,
\eqlabel{action}
\end{equation}
constraints 
\begin{equation}
\a_1=\frac 32\,,\qquad \a_2=\frac 18\,,\qquad \a_3=\frac 16\,.
\eqlabel{ares}
\end{equation}
These values of $\a_i$ agree with those reported in \cite{bk,b1,ls}.

The boundary stress  energy tensor $T^{\mu\nu}$ is obtained from the variation of the full action
\begin{equation}
S_{tot}=S_E^r+S_{GH}+S^{counter}\,,
\eqlabel{stot}
\end{equation} 
with respect to the (Minkowski) boundary metric $\eta_{\mu\nu}=c_{\mu}^{-2}\ h_{\mu\nu}$
\begin{equation}
\begin{split}
T^{\mu\nu}=&2\frac{\dd S_{tot}}{\dd \eta_{\mu\nu}}=\frac{c_1 c_2^3}{c_{\mu}^{-2}}\ \frac{2}{\sqrt{-h}}\ 
\frac{\dd S_{tot}}{\dd h_{\mu\nu}}\bigg|_{r\to \infty}\\
=&\frac{c_1 c_2^3 c_{\mu}^2}{8\pi G_5}\biggl[-\Theta^{\mu\nu}+\Theta h^{\mu\nu}-2\biggl\{
\a_1+\a_2\ \chi^2+\a_3\ \left(\ln \phi\right)^2
\biggr\}\ h^{\mu\nu}+\calo(\chi^4)\biggr]\bigg|_{r\to\infty}\,,
\end{split}
\eqlabel{deft}
\end{equation} 
where 
\begin{equation}
\Theta^{\mu\nu}=\frac 12\left(\nabla^\mu n^\nu+\nabla^\nu n^\mu\right)\,,\qquad \Theta=\tr \Theta^{\mu\nu}\,.
\eqlabel{deftheta}
\end{equation}

\subsection{Asymptotics of the background geometry}

Within the background ansatz \eqref{back}-\eqref{c2def}, it is straightforward to derive from \eqref{ac5} 
equations of motion for $A, H, g, \chi$ and solve these equations perturbatively near the boundary 
$x\to 0_+$, and near the horizon $y=1-x\to 0_+$. These 
equations enjoy exact scaling symmetry 
\begin{equation}
\left(A,\ g,\ H,\ \chi \right)\ \sim\ \left(\Lambda\ A,\ \Lambda\ g,\ H,\ \chi \right)\,,
\eqlabel{scaling}
\end{equation}
for any constant $\Lambda$.

\subsubsection{The boundary $x\to 0_+$ asymptotics}
We find 
\begin{equation}
\begin{split}
A=&a_0+a_1\ x^{1/2}+\left(\frac{a_1^2h_1}{4g_0^2}-2h_1-\frac{1}{72}c_0^2\right)a_1\ x+\calo\left(x^{3/2}\ln x\right)\,,
\end{split}
\eqlabel{ab}
\end{equation}
\begin{equation}
\begin{split}
g=& g_0\left(1+\frac{1}{24}c_0^2\ x^{1/2}+\left(\frac{a_1^2 h_1 c_0^2}{96g_0^2}-\frac{7c_0^4}{1728}
+\frac14+ \frac{h_1 c_0^2}{24}-\frac{a_1^2}{8g_0^2}\right)\ x+\calo\left(x^{3/2}\ln^2 x\right)\right)\,,
\end{split}
\eqlabel{gb}
\end{equation}
\begin{equation}
\begin{split}
H=&1+h_1\ x^{1/2}+\left( \frac{a_1^2 h_1}{4g_0^2}+\frac{c_0^2}{9}\right)h_1\ x+\calo\left(x^{3/2}\ln x\right) \,,
\end{split}
\eqlabel{hb}
\end{equation}
\begin{equation}
\begin{split}
\chi=c_0\ x^{1/4}\ \left(1+\left(c_{10}+\frac{c_0^2}{24}\ \ln x\right) x^{1/2}+\calo\left(x\ln x\right)\right)\,.
\end{split}
\eqlabel{chib}
\end{equation}
The general solution is determined by 6  parameters 
\begin{equation}
\biggl(a_0,\ a_1,\ h_1,\ g_0,\ c_0,\ c_{10} \biggr)\,,
\eqlabel{uvparameters}
\end{equation} 
of which one is the temperature (it can be thought roughly  as a scaling parameter in \eqref{scaling}), 
and another other two $\{a_0,\ c_0\}$ 
are the coefficients of the non-normalizable modes related to the $U(1)_R$ chemical potential $\mu=a_0$ and 
 the mass-deformation scale $M\propto c_0$ respectively. 

The remaining parameters $\{a_1,\ h_1,\ c_{10}\}$ correspond to the charge density $\r$
(conjugate to the chemical potential), and the expectation values of the dimension-2 
$\langle\calo_2\rangle\propto h_1$  and the dimension-3 
$\langle\calo_3\rangle\propto c_0 c_{10}$  operators. To establish precise correspondence, 
we need to introduce a 'universal RG scale' --- a universal (in a sense of 
being independent of any scales in the gauge theory: the temperature, the chemical potential, 
and the mass-deformation scale) radial coordinate:
\begin{equation}
\hr\equiv c_2^{-1}=\frac{3^{1/2}h_1^{1/2}}{g_0}\ x^{1/4}\ \left(1+\left(
\frac{a_1^2h_1}{8g_0^2}+\frac{c_0^2}{72}\right) x^{1/2}+\calo\left(x\ln x\right)\right)\,.
\eqlabel{runiver}
\end{equation}   
The precise  values (up to a $c$-number normalization) 
of various mass scales and  operator expectation values (VEV's) can then be obtained
as the coefficients of the non-normalizable/normalizable modes of the corresponding gravitational fields 
with a radial dependence given by \eqref{runiver}:
\begin{equation}
\begin{split}
&M=c_0\ \frac{g_0}{3^{1/2} h_1^{1/2}}\,,\qquad \r=-\frac{1}{8\pi G_5}\ a_1\ 
\left(\frac{g_0}{3^{1/2}h_1^{1/2}}\right)^2\,,\qquad 
\langle\calo_2\rangle=h_1\ \left(\frac{g_0}{3^{1/2}h_1^{1/2}}\right)^2\,,
\\
&\langle\calo_3\rangle=c_0\ \left(\frac{g_0}{3^{1/2} h_1^{1/2}}\right)^3\biggl(c_{10}- 
\frac {a_1^2 h_1}{8g_0^2}\biggr)
+\calo\left(g_0^3 c_0^3\right)\,.
\end{split}
\eqlabel{scalesuv}
\end{equation} 
The factor of $(-8\pi G_5)^{-1}$ in the definition of $\rho$ is inserted to insure consistency of the 
thermodynamics of the dual gauge theory.  Notice that  the expectation value $\langle\calo_3\rangle$
is determined only up to order $\calo(c_0^2)$ --- this is related to the fact that the effective action 
\eqref{ac5} is defined only to this order as well.

\subsubsection{The horizon $y=1-x\to 0_+$ asymptotics}
We find
\begin{equation}
A=a_2^h\ y^2+\calo(y^4)\,,
\eqlabel{aah}
\end{equation}
\begin{equation}
g=g_0^h+g_2^h\ y^2+\calo(y^4)\,,
\eqlabel{ggh}
\end{equation}
\begin{equation}
\begin{split}
H=&h_0^h-\frac{(4 (h_0^h)^6 (a_2^h)^2+(h_0^h)^4 (a_2^h)^2 (c_0^h)^2+4 (h_0^h)^3 (a_2^h)^2+8 g_2^h g_0^h) h_0^h 
((h_0^h)^3-1)}{3(g_0^h)^2 (h_0^h (c_0^h)^2+4)}\ y^2
\\
&+\calo(y^4)\,,
\end{split}
\eqlabel{hhh}
\end{equation}
\begin{equation}
\chi=c_0^h-\frac{3 h_0^h c_0^h ((h_0^h)^6 (a_2^h)^2+2 g_2^h g_0^h)}{(g_0^h)^2 (h_0^h (c_0^h)^2+4)}\ y^2
+\calo(y^4)\,.
\eqlabel{cch}
\end{equation}
The most general solution is determined by 5 parameters
\begin{equation}
\biggl(a_2^h,\ g_0^h,\ g_2^h,\ h_0^h,\ c_0^h\biggr)\,.
\eqlabel{irparameters}
\end{equation}

Note that once the temperature, the chemical potential and the deformation scale are fixed, 
we have $5+6-3=8$ parameters, which is the correct number to uniquely determine the solution.

\subsection{Thermodynamics of the background geometry}
Given \eqref{scalesuv}, the thermodynamics of the mass-deformed theory must be studied under the constraint
\begin{equation}
c_0=\frac{3^{1/2} h_1^{1/2}}{g_0}\ M\,,
\eqlabel{scale}
\end{equation}
for a fixed $M$.

All the thermodynamic quantities are given to order $\calo(c_0^2)$ or $\calo((c_0^h)^2)$.

The temperature $T$ and the chemical potential $\mu$  of the background geometry is given by
\begin{equation}
(2\pi T)^2=\frac{(h_0^h (c_0^h)^2+4)(g_0^h)^4}{4((h_0^h)^3-1)((h_0^h)^6(a_2^h)^2+2g_2^hg_0^h)}\,,\qquad \mu=a_0\,.
\eqlabel{tmu}
\end{equation}
The entropy density $s$ is 
\begin{equation}
s=\frac{\cala_{hor}}{4 G_5}=\frac{c_2^3}{4G_5}\bigg|_{y\to 0_+}=\frac{1}{8\pi G_5}\ \frac{2\pi(g_0^h)^3(h_0^h)^{3/2}}{((h_0^h)^3-1)^{3/2}}\,.
\eqlabel{s}
\end{equation}
The regularized Euclidean action \eqref{action} (up to the space-time volume factor) has to be identified 
with the Gibbs free energy density $\om$ 
\begin{equation}
\begin{split}
\om=&\left(\int_{\del\calm_5} d^4\xi\right)^{-1}\ \cali_E
=\frac{1}{8\pi G_5}\biggl(-\frac{g_0^4}{9h_1^2}+\frac{g_0^2 (a_1^2 h_1-8 g_0^2 c_{10}) c_0^2}{288h_1^2}\biggr)\,.
\end{split}
\eqlabel{om}
\end{equation}
From \eqref{deft}, the energy density $\e$ and the pressure $P$ are given by
\begin{equation}
\begin{split}
\e=&\frac{1}{8\pi G_5}\ \biggl(\frac{g_0^4}{3h_1^2}+\frac{g_0^2 (a_1^2 h_1-8 g_0^2 c_{10}) c_0^2}{288h_1^2}\biggr)\,,\\
P=&\frac{1}{8\pi G_5}\ \biggl( \frac{g_0^4}{9h_1^2}-\frac{g_0^2 (a_1^2 h_1-8 g_0^2 c_{10}) c_0^2}{288h_1^2}\biggr)\,.
\end{split}
\eqlabel{ep}
\end{equation}
Notice that $\om=-P$, as required by the Minkowski space-time thermodynamics.  
Finally, from \eqref{scalesuv}, the charge density is given by 
\begin{equation}
\r=-\frac{1}{8\pi G_5}\ \frac{a_1g_0^2}{3h_1} \,.
\eqlabel{rhoa}
\end{equation}

While not automatically satisfied, we expect the basic thermodynamic relations 
\begin{equation}
\om=\e-T s -\mu \r\,,\qquad d\e=T\ ds+\mu\ d\r\,,\qquad dP=s\ dT+\rho\ d\mu\,,
\eqlabel{basict}
\end{equation}
to hold.

\subsection{Thermodynamics to order $\calo(\l^0)$}\label{therm0}

The background geometry to this order is given by \eqref{zeroback}.
Thus, we find (we do not need $c_{10}$)
\begin{equation}
\biggl(a_0,\ a_1,\ h_1,\ g_0,\ c_0\biggr)=\biggl(
\frac{\b}{\sqrt{1+\k}},\ -2^{1/2}\b,\ \frac{2^{1/2}\k}{3\sqrt{1+\k}},\ \b,\ 
0   \biggr)\,,
\eqlabel{uv0}
\end{equation}
and 
\begin{equation}
\biggl(a_2^h,\ g_0^h,\ g_2^h,\ h_0^h,\ c_0^h\biggr)=\biggl(\frac{\b}{\sqrt{1+\k}(\k+2)},\ \b,\ 0,\ (1+\k)^{1/3},\ 0\biggr)\,,
\eqlabel{ir0}
\end{equation}
leading to (we use \eqref{g5})
\begin{equation}
\begin{split}
&s=\frac{4 \pi^2 (1+\k)^2 T^3 N^2}{(\k+2)^3}\,,\qquad \e=3 P=\frac{6 N^2 T^4 (1+\k)^3 \pi^2}{(\k+2)^4}
\,,\\
&\r=\frac{2\pi (1+\k)^2 \k^{1/2} T^3 N^2}{(\k+2)^3}\,,\qquad \frac{2\pi T}{\mu}=\sqrt{\k}+\frac {2}{\sqrt{\k}}\,.
\end{split}
\eqlabel{thermo0}
\end{equation}
It is straightforward to verify the basic thermodynamic 
relations \eqref{basict}.

From \eqref{thermo0} we see that $\frac{T}{\mu}$ achieves a minimum at $\k=\k_c=2$, corresponding to the critical temperature 
$T_c=\sqrt{2}\mu/\pi$ and the critical chemical potential $\mu_c=\pi T/\sqrt{2}$. Introducing 
\begin{equation}
t=\frac{T}{T_c}-1\,,\qquad \bm=1-\frac{\mu}{\mu_c}\qquad \Longrightarrow\qquad  \bm=\frac{t}{t+1}\,,
\eqlabel{deftred}
\end{equation}
we find 
\begin{equation}
\begin{split}
\om=&\om_{\pm}(\mu,t)=-\frac{27N^2\mu^4}{32\pi^2}\left(1+\frac 83\ t\mp\frac{16\sqrt{2}}{27}\ t^{3/2}+\frac{68}{27}\ t^2+\calo\left(t^{5/2}\right)\right)\,,\\
=&\om_{\pm}(T,\bm)=-\frac{27N^2 T^4\pi^2}{128}\left(1-\frac 43\ \bm\mp\frac{16\sqrt{2}}{27}\ \bm^{3/2}+\frac{14}{27}\ \bm^2
+\calo\left(\bm^{5/2}\right)\right)\,,
\end{split}
\eqlabel{omcrit}
\end{equation}
\begin{equation}
\begin{split}
\k=&\k_{\pm}(t)=2\pm4\sqrt{2}\ t^{1/2} +8\ t\pm5\sqrt{2}\ t^{3/2}+4\ t^2+\calo\left(t^{5/2}\right)\\
=&\k_{\pm}(\bm)=2\pm 4\sqrt{2}\ \bm^{1/2} +8\ \bm\pm7\sqrt{2}\ \bm^{3/2}+12\ \bm^2+\calo\left(\bm^{5/2}\right)\,,
\end{split}
\eqlabel{kappacrit}
\end{equation}
where the signs in both expressions correlate. Thus for a given temperature 
$t$ there are two thermodynamic phases of the system, 
with $\om_-$ being the stable one. Specific heat of the $\om_+$ phase is 
negative, and thus this phase is thermodynamically unstable. In \cite{b5} it was argued 
that in the absence of the chemical potentials in holographic examples of 
gauge theory/string theory correspondence the thermodynamic instabilities 
show up as instabilities in the propagation of  sound waves.  
In RN plasma of interest here there are no instabilities in the sound channel; 
rather, we find that in the $\om_+$ phase the two-point correlation function 
of the order parameter (the charge density) in the vicinity of the critical point 
oscillates, instead of exponentially decaying. Additionally, the
dynamical relaxation time in the $\om_+$ phase is {\it negative}, which explicitly 
demonstrates the instability of this phase (see section 4).   

For $\om_-$ phase we find:
\begin{equation}
\begin{split}
C=&T\left(\frac{\del s}{\del T}\right)\bigg|_{\mu}\propto -\frac{\del^2\om_-(\mu,t)}{\del t^2}\propto +t^{-1/2}
\,,\\
\r=&-\left(\frac{\del \om_-}{\del \mu}\right)\bigg|_T\propto -\frac{\om_-(T,\bm)}{\del \bm}
\propto +\bm^{1/2}\propto +t^{1/2}\,,\\
\chi_T=&\left(\frac{\del\r}{\del\mu}\right)\bigg|_T\propto -\frac{\del^2\om_-(T,\bm)}{\del \bm^2}\propto +\bm^{-1/2}\propto + t^{-1/2}\,,
\end{split}
\eqlabel{cchi}
\end{equation}
in the vicinity of the phase transition.

Recall that in the classic theory of static critical phenomena one 
introduces critical exponents 
\begin{equation}
\left(\a\,,\b\,,\gamma\,,\delta\,,\nu\,,\eta\right)\,.
\eqlabel{statcrit}
\end{equation}
Once we identify the charge density $\r$ of the  RN plasma with the order parameter, and 
the chemical potential $\mu$ with the external (control) parameter we can read 
from \eqref{cchi} the following 4 critical exponents
\begin{equation}
\left(\a\,,\b\,,\gamma\,,\delta\right)\ =\ \left(\frac 12\,, \frac 12\,, \frac 12\,, 2
\right)\,.
\eqlabel{statcritn4}
\end{equation}
Under a single scale hypothesis in a continuous critical phenomena only 
two static critical exponents are independent --- there are 4 scaling relations:
\begin{equation}
\a+2\b+\gamma=2\,,\qquad \gamma=\b(\delta-1)=\nu(2-\eta)\,,\qquad 2-\a=\nu p\,,
\eqlabel{scalingn4}
\end{equation}
where $p$ is the number of spatial dimensions of a critical system.
Notice that given \eqref{statcritn4}, the first two scaling relations in \eqref{scalingn4}
are satisfied. One can use the rest of the scaling relations to determine the remaining 
static critical exponents:
\begin{equation}
\nu\bigg|_{relation}=\frac 12\,,\qquad \eta\bigg|_{relation}=1\,.
\eqlabel{nueta}
\end{equation}
As we will see in section 4, the values of $(\nu,\eta)$ as in \eqref{nueta} are incorrect. 
We point only that the value of $\eta$ in \eqref{nueta} already raises suspicion:
we consider here the critical phenomena in RN plasma in the strict 't Hooft limit, \ie, 
as $N\to \infty$ --- in this limit we expect a mean-field criticality, leading to 
\begin{equation}
\eta\bigg|_{mean-field}=0\,.
\eqlabel{etaexpe}
\end{equation}

\subsection{Thermodynamics to order $\calo(\l^2)$}
Since we are going to be interested in the deformation of the critical phenomena in \eqref{omcrit} at order $\calo(\l^2)$
(or more precisely $\calo(M^2)$, see \eqref{scale}) we have to solve \eqref{eq4} and \eqref{eq3} in the vicinity of $\k=2$. 
Thus, we further expand 
\begin{equation}
\begin{split}
&g_2=\b \sum_{n=0}^\infty (\k-2)^n\ g_2^{(n)}\,,\qquad  H_2=\sum_{n=0}^\infty (\k-2)^n\ H_2^{(n)}\,,\qquad
A_2=\b \sum_{n=0}^\infty (\k-2)^n\ A_2^{(n)}\,,\\
&\c_1=\sum_{n=0}^\infty (\k-2)^n\ \c_1^{(n)} \,.
\end{split}
\eqlabel{critexpansion}
\end{equation}
Clearly, $\k=2$ is not a singular point of \eqref{eq4} and \eqref{eq3}, thus \eqref{critexpansion} leads to a series of smooth 
ODE's for $\{g_2^{(n)},\ H_2^{(n)},\ A_2^{(n)},\ \c_1^{(n)}\}$. These 
ODE's must be solved with the following boundary conditions:
\begin{equation}
\begin{split}
&x\to 0_+:\qquad g_2^{(n)}\to 0\,,\qquad H_2^{(n)}\to 0\,,\qquad A_2^{(n)}\to 0\,,
\qquad \c_1^{(n)}\to \delta^{n}_0\ x^{1/4}\,,\\
&x\to 1_-:\qquad   g_2^{(n)}\to {\rm const}\,,\qquad H_2^{(n)}\to {\rm const}\,,\qquad A_2^{(n)}\to 0\,,
\qquad \c_1^{(n)}\to {\rm const}\,.\\
\end{split}
\eqlabel{boundconditions}
\end{equation}
The normalization of the non-normalizable modes of $A_2^{(n)}$ and $\chi_1$ near the boundary implies from \eqref{tmu}
and \eqref{scalesuv}
that 
\begin{equation}
\l=\frac{(\k+2)}{2^{3/4}(1+\k)^{3/4}\pi}\ \frac MT\left(1+\calo\left(\frac{M^2}{T^2}\right)\right)\,.
\eqlabel{ladef}
\end{equation}

It is straightforward to construct asymptotic solutions for $\{g_2^{(n)},\ H_2^{(n)},\ A_2^{(n)},\ \c_1^{(n)}\}$,
subject to the boundary conditions \eqref{boundconditions} --- with obvious modifications they take the form 
of \eqref{ab}-\eqref{chib} and \eqref{aah}-\eqref{cch}. Thus we find:
\nxt for UV parameters \eqref{uvparameters}:
\begin{equation}
\begin{split}
&a_0=\frac{\b}{\sqrt{1+\k}}\,,\qquad a_1=-\b\sqrt{2}+\l^2 \b\sum_{n=0}^\infty (\k-2)^n\ a_1^{(n)} \,,\\
&h_1=\frac{\k\sqrt{2}}{3\sqrt{1+\k}}+\l^2\sum_{n=0}^\infty 
(\k-2)^n\ h_1^{(n)}\,,\qquad
g_0=\b\,,\qquad c_0=\l\,,\\
&c_{10}=\sum_{n=0}^\infty (\k-2)^n\ c_{10}^{(n)}\,,
\end{split}
\eqlabel{uvexpand}
\end{equation}
\nxt for IR parameters \eqref{irparameters}:
\begin{equation}
\begin{split}
&a_2^h=\frac{\b}{(\k+2)\sqrt{1+\k}}+\l^2\b\sum_{n=0}^\infty (\k-2)^n\ a_2^{h(n)}\,,\qquad
g_0^h=\b+\l^2\b\sum_{n=0}^\infty (\k-2)^n\ g_0^{h(n)}\,,\\
&g_2^h=\l^2\b\sum_{n=0}^\infty (\k-2)^n\ g_2^{h(n)}\,,\qquad h_0^h=(1+\k)^{1/3}+\l^2\sum_{n=0}^\infty (\k-2)^n\ h_0^{h(n)}\,,\\
&c_0^h=\l\sum_{n=0}^\infty (\k-2)^n\ c_0^{h(n)}\,,
\end{split}
\eqlabel{irexpand}
\end{equation}
where $\l$ is given by \eqref{ladef}. 

Notice that at each order $(n)$ there are eight parameters 
\begin{equation}
\biggl(\ 
a_1^{(n)}\,,\ h_1^{(n)}\,,\ c_{10}^{(n)}\,,\ a_2^{h(n)}\,,\ g_0^{h(n)}\,,\ g_2^{h(n)}\,,\ h_0^{h(n)}\,,\ c_0^{h(n)}
\ \biggr)\,,
\eqlabel{totalp}
\end{equation}
which are uniquely determined by solving ODE's for $\{g_2^{(n)},\ H_2^{(n)},\ A_2^{(n)},\ \c_1^{(n)}\}$, subject to the boundary
conditions \eqref{boundconditions}.

For the first two orders --- $n=0,1$ ---  we collected these coefficients in Table~\ref{tab1}.

\begin{table}
\centerline{
\\
\begin{tabular}
{||c||c|c|c||}
	\hline
\textbf{\em n}  &  $0$    &   $1$ \\
\hline
\hline
$a_1^{(n)}$ &  -0.1655555689119159 & -0.013078142827910066\\
\hline
$h_1^{(n)}$ & 0.11928166889409134 & 0.018530943268730934
\\
\hline
$c_{10}^{(n)}$ & -0.23071409088829617 &-0.01240068268619082
\\
\hline
$a_2^{h(n)}$ & -0.03250808756595229&0.005503802347302805
\\
\hline
$g_0^{h(n)}$ & 0.021412846345792092&0.0015030136731377015
\\
\hline
$g_2^{h(n)}$ & - 0.007908684866787086&-0.0009615292197374708
\\
\hline
$h_0^{h(n)}$ & 0.1485225060823788 &0.042145056457606175
\\
\hline
$c_0^{h(n)}$ &0.7464562054847809 &-0.0129139123278239
\\
\hline
\end{tabular}
}
\caption{Coefficients of the normalizable modes of the background geometry.}
\label{tab1}
\end{table}

Given \eqref{ladef}, \eqref{uvexpand} and \eqref{irexpand} we can compute following \eqref{ep} and \eqref{rhoa}
thermodynamic potentials as a series in $(\k-2)$ and to order $\calo\left(\frac{M^2}{T^2}\right)$.
Recall that the basic thermodynamic relations \eqref{basict} are not automatically satisfied. 
Rather, they impose algebraic constraints on  \eqref{totalp}. We find:
\nxt at order $n=0$:
\begin{equation}
\begin{split}
&0=7 \sqrt{2}\ 3^{2/3}\ h_0^{h(0)}-\frac{15}{2} \sqrt{3}\ h_1^{(0)}
-\frac 14 \sqrt{2}\ 3^{1/3}\ (c_{0}^{h(0)})^2+\frac{32}{3} \sqrt{2}\ g_2^{h(0)}
-10 \sqrt{2}\ g_0^{h(0)}\\
&+8  \sqrt{6}\ a_2^{h(0)}+a_1^{(0)}\,,
\end{split}
\eqlabel{const0}
\end{equation}
\begin{equation}
\begin{split}
&0=-12 \sqrt{3}\ h_1^{(0)}-\frac 34 \sqrt{2}\ 3^{1/3}\ (c_0^{h(0)})^2+14 \sqrt{2}\ 3^{2/3}\ h_0^{h(0)}
+\frac{1}{24} \sqrt{3}+32 \sqrt{2}\ g_2^{h(0)}\\
&-18 \sqrt{2}\ g_0^{h(0)}
-\frac38 \sqrt{2}\ c_{10}^{(0)}+a_1^{(0)}+24 \sqrt{6}\ a_2^{h(0)}\,,
\end{split}
\eqlabel{fk0}
\end{equation}
\begin{equation}
\begin{split}
&0=-\frac{3}{16} \sqrt{2}\ c_{10}^{(1)}+\frac{1}{16} \sqrt{2}\ c_{10}^{(0)}+4 \sqrt{3}\ h_1^{(0)}
-\frac{21}{4} \sqrt{3}\ h_1^{(1)}-\frac{14}{3} \sqrt{6}\ a_2^{h(0)}
-\frac{73}{18} \sqrt{2}\ 3^{2/3}\ h_0^{h(0)}\\
&+\frac 72 \sqrt{2}\ 3^{2/3}\ h_0^{h(1)}+\frac{23}{6} \sqrt{2}\ g_0^{h(0)}-6 \sqrt{2}\ 
g_0^{h(1)}-\frac{56}{9} \sqrt{2}\ g_2^{h(0)}+\frac{7}{48} \sqrt{2}\ 3^{1/3}\ (c_0^{h(0)})^2+a_1^{(1)}\,,
\end{split}
\eqlabel{fm0}
\end{equation}
\nxt at order $n=1$:
\begin{equation}
\begin{split}
&0=-\frac 16\ a_1^{(0)}+\frac{21}{4} \sqrt{3}\ h_1^{(0)}-\frac 83 \sqrt{2}\ g_2^{h(0)}+\frac{11}{144} \sqrt{2}\ 3^{1/3}\ (c_0^{h(0)})^2
-\frac{169}{36} \sqrt{2}\ 3^{2/3}\ h_0^{h(0)}\\
&+\frac{25}{6} \sqrt{2}\ g_0^{h(0)}+\frac{32}{3} \sqrt{2}\ g_2^{h(1)}+8 \sqrt{6}\ a_2^{h(1)}-10 \sqrt{2}\ g_0^{h(1)}+7 \sqrt{2}\ 3^{2/3}\ h_0^{h(1)}
-\frac{15}{2} \sqrt{3}\ h_1^{(1)}\\
&-\frac 12 \sqrt{2}\ 3^{1/3}\ c_0^{h(0)} c_0^{h(1)}+a_1^{(1)}\,,
\end{split}
\eqlabel{const1}
\end{equation}
\begin{equation}
\begin{split}
&0=14 \sqrt{2}\ 3^{2/3}\ h_0^{h(1)}-\frac 32 \sqrt{2}\ 3^{1/3}\ c_0^{h(0)} c_0^{h(1)}+\frac{11}{48} \sqrt{2}\ 3^{1/3}\ (c_0^{h(0)})^2
-\frac{80}{9} \sqrt{2}\ 3^{2/3}\ h_0^{h(0)}\\
&-8 \sqrt{2}\ g_2^{h(0)}+\frac{15}{2} \sqrt{2}\ g_0^{h(0)}+\frac 18 \sqrt{2}\ c_{10}^{(0)}+32 \sqrt{2}\
 g_2^{h(1)}
-18 \sqrt{2}\ g_0^{h(1)}-\frac 38 \sqrt{2}\ c_{10}^{(1)}-\frac 16\ a_1^{(0)}\\
&+a_1^{(1)}+24 \sqrt{6}\ a_2^{h(1)}+\frac{33}{4} \sqrt{3}\
 h_1^{(0)}-12 \sqrt{3}\ h_1^{(1)}\,,
\end{split}
\eqlabel{fk1}
\end{equation}
\begin{equation}
\begin{split}
&0=\frac{7}{48} \sqrt{2}\ 3^{1/3}\ c_0^{h(0)} c_0^{h(1)}+\frac{19}{6} \sqrt{2}\ g_0^{h(1)}-\frac{1}{12}\ a_1^{(1)}+a_1^{(2)}+\frac{23}{18} \sqrt{6}\ a_2^{h(0)}
-\frac 73 \sqrt{6}\ a_2^{h(1)}\\
&-\frac{61}{24} \sqrt{3}\ h_1^{(0)}+\frac{31}{8} \sqrt{3}\ h_1^{(1)}-\frac{21}{4} \sqrt{3}\ h_1^{(2)}+\frac{67}{27}
 \sqrt{2}\ g_2^{h(0)}-\frac{29}{16} \sqrt{2}\ g_0^{h(0)}\\
&-\frac{215}{3456} \sqrt{2}\ 3^{1/3}\ (c_0^{h(0)})^2-\frac{28}{9} \sqrt{2}\ g_2^{h(1)}+\frac{3235}{1296} \sqrt{2}\ 3^{2/3}\ h_0^{h(0)}
+\frac{1}{16} \sqrt{2}\ c_{10}^{(1)}-\frac{11}{384} \sqrt{2}\ c_{10}^{(0)}\\
&-\frac{3}{16} \sqrt{2} c_{10}^{(2)}-\frac{479}{144} \sqrt{2}\ 3^{2/3}\ h_0^{h(1)}+\frac 72 \sqrt{2}\ 3^{2/3}\ h_0^{h(2)}-6 \sqrt{2}\ g_0^{h(2)}\,.
\end{split}
\eqlabel{fm1}
\end{equation}

Given data in Table~\ref{tab1} we can verify \eqref{const0}-\eqref{fk0} to agree with a relative accuracy of $\sim 10^{-10}$,
and \eqref{fm0}-\eqref{fk1} with a relative accuracy of $\sim 10^{-7}$.

We are now in position to evaluate the free energy near the critical point. 
First of all, notice that $\k_c$$, T_c$ or $\mu_c$ will receive $\calo(\l^2)$  corrections:
\begin{equation}
\begin{split}
\k_c=&2+\frac{M^2}{\pi^2 T^2}\biggl(\frac{512\sqrt{6}}{81}\ g_2^{h(0)}+\frac{320\sqrt{2}}{9}\ a_2^{h(0)}
-\frac{184\sqrt{2}}{27} 3^{1/6}\  h_0^{h(0)}-\frac{8\sqrt{2}}{81} 3^{5/6}\ \left(c_0^{h(0)}\right)^2\\
&+\frac{112\sqrt{2}}{3} 3^{1/6}\ h_0^{h(1)}+\frac{256\sqrt{2}}{3}\  a_2^{h(1)}
+\frac{1024\sqrt{6}}{27}\ g_2^{h(1)}-\frac{128\sqrt{6}}{9}\  g_0^{h(1)}\\
&-\frac{16\sqrt{2}}{9} 3^{5/6} 
c_0^{h(0)} c_0^{h(1)}\biggr)+\calo\left(\frac{M^4}{T^4}\right)\,,
\end{split}
\eqlabel{kc}
\end{equation}
\begin{equation}
\begin{split}
T_c=&\frac{\sqrt{2}}{\pi}\mu\biggl(1+\frac{M^2}{\mu^2}\biggl(\frac 49 \sqrt{6}\ g_0^{h(0)}-\frac{32}{27} \sqrt{6}\ g_2^{h(0)}
-\frac 83 \sqrt{2}\ a_2^{h(0)}-\frac 76 \sqrt{2}\ 3^{1/6}\ h_0^{h(0)}\\
&+\frac{1}{36} \sqrt{2}\ 3^{5/6} \left(c_0^{h(0)}\right)^2\biggr)+\calo\left(\frac{M^4}{\mu^4}\right)\biggr)\,,
\end{split}
\eqlabel{tcl}
\end{equation}
\begin{equation}
\begin{split}
\mu_c=&\frac{{\pi}}{\sqrt{2}}T\biggl(1+\frac{M^2}{\pi^2T^2}\biggl(-\frac 89 \sqrt{6}\ g_0^{h(0)}+\frac{64}{27} \sqrt{6}\ g_2^{h(0)}\
+\frac{16}{3} \sqrt{2}\ a_2^{h(0)}+\frac 73 \sqrt{2}\ 3^{1/6}\ h_0^{h(0)}\\
&-\frac{1}{18} \sqrt{2}\ 3^{5/6} \left(c_0^{h(0)}\right)^2\biggr)+\calo\left(\frac{M^4}{T^4}\right)\biggr)\,,
\end{split}
\eqlabel{mucl}
\end{equation}
with the mass deformation  {\it lowering}\footnote{We used results of Table~\ref{tab1}.} the critical temperature for a fixed chemical potential.
Once again, introducing 
\begin{equation}
t=\frac{T}{T_c}-1\,,\qquad \bm=1-\frac{\mu}{\mu_c}\,,
\eqlabel{defag}
\end{equation}
we find
\begin{equation}
\begin{split}
\om_{\pm}(\mu,t)=&-\frac{27N^2\mu^4}{32\pi^2}\left(1+s_t^0\ \frac{M^2}{\mu^2}\right)\biggl(1\pm s_t^1\  \frac{M^2}{\mu^2}\ t^{1/2}
+\frac 83\left(1+s_t^2\ \frac{M^2}{\mu^2}\right)\ t\\
&\mp\frac{16\sqrt{2}}{27}\left(1+s_t^3\ \frac{M^2}{\mu^2}\right)\ t^{3/2}+\dots+\calo\left(\frac{M^4}{\mu^4}\right)
\biggr)\,,
\end{split}
\eqlabel{omtm}
\end{equation}
\begin{equation}
\begin{split}
\om_{\pm}(T,\bm)=&-\frac{27N^2 T^4\pi^2}{128}\left(1+s_{\bm}^0\ \frac{M^2}{\pi^2T^2}\right)\biggl(1\pm s_{\bm}^1\  \frac{M^2}{\pi^2T^2}\ t^{1/2}
-\frac 43\left(1+s_{\bm}^2\ \frac{M^2}{\pi^2T^2}\right)\ \bm\\
&\mp\frac{16\sqrt{2}}{27}\left(1+s_{\bm}^3\ \frac{M^2}{\pi^2T^2}\right)\ \bm^{3/2}+\cdots+
+\calo\left(\frac{M^4}{T^4}\right)\biggr)\,.
\end{split}
\eqlabel{ommum}
\end{equation}
For the first two corrections to \eqref{omcrit} we have
\begin{equation}
\begin{split}
s_t^0=&-2\ h_1^{(0)}-\frac{1}{5}+\frac{1}{18} \sqrt{6}\ c_{10}^{(0)}\,,
\end{split}
\eqlabel{st0}
\end{equation}
\begin{equation}
\begin{split}
s_t^1=&-8 \sqrt{2}\ h_1^{(1)} +\frac 83 \sqrt{2}\ h_1^{(0)}-\frac{16}{243}\ 3^{5/6} \left(c_0^{h(0)}\right)^2-\frac{32}{27}\ 3^{5/6}\
 c_0^{h(0)} c_0^{h(1)}+\frac{224}{9}\ 3^{1/6}\ h_0^{h(1)}\\
&-\frac{368}{81}\ 3^{1/6}\ h_0^{h(0)}-\frac{256}{27} \sqrt{3}\ g_0^{h(1)}+\frac{1024}{243} \sqrt{3}\ g_2^{h(0)}+\frac{2048}{81} \sqrt{3}\ 
g_2^{h(1)}+\frac{640}{27}\ a_2^{h(0)}\\
&+\frac{512}{9}\ a_2^{h(1)}-\frac{2}{81} \sqrt{2}+\frac 49 \sqrt{3}\ c_{10}^{(1)}\,,
\end{split}
\eqlabel{st1}
\end{equation}
\begin{equation}
\begin{split}
s_{\bm}^0=&-4\ h_1^{(0)}+\frac 19 \sqrt{6}  c_{10}^{(0)}-\frac{1}{27}-\frac{32}{9} \sqrt{6}\ g_0^{h(0)}+\frac{256}{27} \sqrt{6} \ g_2^{h(0)}
+\frac{28}{3}\ 3^{1/6}\ \sqrt{2}\ h_0^{h(0)}\\
&+\frac{64}{3} \sqrt{2}\ a_2^{h(0)}-\frac 29\ 3^{5/6}\ \sqrt{2} \left(c_0^{h(0)}\right)^2\,,
\end{split}
\eqlabel{sm0}
\end{equation}
\begin{equation}
\begin{split}
s_{\bm}^1=&-16  \sqrt{2}\ h_1^{(1)}
+\frac{16}{3} \sqrt{2}\ h_1^{(0)}-\frac{32}{243}\ 3^{5/6} \left(c_0^{h(0)}\right)^2-\frac{64}{27}\ 3^{5/6}\ c_0^{h(0)} c_0^{h(1)}
+\frac{448}{9}\ 3^{1/6}\ h_0^{h(1)}\\
&-\frac{736}{81}\ 3^{1/6}\ h_0^{h(0)}-\frac{512}{27} \sqrt{3}\ g_0^{h(1)}+\frac{2048}{243} \sqrt{3}\
 g_2^{h(0)}+\frac{4096}{81} \sqrt{3}\ g_2^{h(1)}+\frac{1280}{27}\ a_2^{h(0)}\\
&+\frac{1024}{9}\ a_2^{h(1)}-\frac{4}{81} \sqrt{2}+\frac 89
 \sqrt{3}\ c_{10}^{(1)}\,.
\end{split}
\eqlabel{sm1}
\end{equation}

It is straightforward to check, that given constraints from the basic thermodynamic relations \eqref{const0}-\eqref{fk1} 
\begin{equation}
s_t^1=s_{\bm}^1=0\,,
\eqlabel{van1}
\end{equation}
which guarantees that the mass-deformed theory 
has the same static critical exponents $(\a,\b,\gamma,\delta)$
as in \eqref{statcritn4}.

\section{Fluctuations in charged plasma and its holographic 
dual}\label{fluctuations}

\subsection{Hydrodynamic modes in charged plasma}
In this section we consider the propagation of hydrodynamic modes in four-dimensional charged relativistic plasma in the absence of background 
electro-magnetic fields. 

The first order hydrodynamic equations of motion
in Minkowski space-time are simply the conservation laws for the stress-energy tensor and the $U(1)$ current:
\begin{equation}
\del_\nu T^{\mu\nu}=0\,,\qquad \del_\mu J^\mu=0\,.
\eqlabel{2.10}
\end{equation}
One can do the standard decomposition of the stress tensor,
\begin{equation}
T^{\mu \nu} =\epsilon u^{\mu}u^{\nu}+P \Delta^{\mu \nu} +\Pi^{\mu \nu}\,, 
\eqlabel{2.11}
\end{equation}
where 
\begin{equation}
\Delta^{\mu \nu}=\eta^{\mu \nu}+u^{\mu}u^{\nu}\,,\;\;\; \Pi^{\mu}_{\;\;\nu} u^\nu = 0\;,
\eqlabel{2.12}
\end{equation}
and $\epsilon$ and $P$ are the energy density and the pressure respectively.  
The dissipative term $\Pi^{\mu \nu}$ is given by
\begin{equation}
\Pi^{\mu \nu}=-\eta \sigma^{\mu \nu} -\zeta \Delta^{\mu \nu} (\partial_{\alpha} u^{\alpha})\,, 
\eqlabel{2.14}
\end{equation}
where 
\begin{equation}
\sigma^{\mu \nu}=\Delta^{\mu \alpha} \Delta^{\nu \beta}
(\partial_{\alpha}u_{\beta}+\partial_{\beta}\partial_{\alpha})
-\frac 23 \Delta^{\mu \nu}\Delta^{\alpha \beta}(\partial_{\gamma}u^{\gamma})\,,
\eqlabel{2.15}
\end{equation}
and $\eta$ and $\zeta$ are the shear and the bulk viscosities. 
Note that $\Pi^{\mu \nu}$ is, by definition, zero at local equilibrium. 
The current $J^{\mu}$  is given by 
\begin{equation}
J^{\mu}=\rho u^{\mu} +\nu^{\mu}\,, 
\eqlabel{2.16}
\end{equation}
where $\nu^{\mu}$ is the dissipative part satisfying $u^{\mu}\nu_{\mu}=0$: 
\begin{equation}
\nu^{\mu}=\sigma_{Q}\Delta^{\mu \nu}\left(-\partial_{\nu}\mu
+\frac{\mu}{T}\partial_{\nu}T\right)\,. 
\eqlabel{2.17}
\end{equation}
In this expression, $T$ is the temperature, $\mu$ is the chemical potential, 
and $\sigma_Q$ is the conductivity coefficient. 
We would like to study fluctuations around the equilibrium state in which 
\begin{equation}
u^{\mu}=(1, 0, 0,0)\,, \quad T={\rm const.}\,, \quad \mu={\rm const.}\,.
\eqlabel{2.18}
\end{equation}
As an independent set of variables we will choose the three spatial components 
of the velocity $\delta u^1=\delta u^x,\ \delta u^2=\delta u^y,\ \delta u^3=\delta u^z$, 
as well as $\delta T$ and $\delta \mu$. As usual, all perturbations 
are of the plane-wave form $exp(-i \omega t+i q z)$. We find that the relevant 
fluctuations of $T^{\mu \nu}$ are
\begin{equation}
\begin{split}
& \delta T^{t t} =\delta \epsilon = 
\left(\frac{\partial \epsilon}{\partial \mu}\right)_{T}
\delta \mu 
+\left(\frac{\partial \epsilon}{\partial T}\right)_{\mu}
\delta T\,,  \\
& \delta T^{t i}=(\e +P)\delta u^i\,, \\
&\delta T^{x z}=-\eta \partial_z \delta u^x\,,\qquad \delta T^{y z}=-\eta \partial_z \delta u^y\,,  \\
&\delta T^{zz}=\delta P- \left(\frac 43\eta +\zeta\right)\partial_z \delta u^z=
\left( \frac{\partial P}{\partial \mu}\right)_{T}
\delta \mu+
\left( \frac{\partial P}{\partial T}\right)_{\mu}
\delta T 
-\left(\frac 43 \eta+\zeta\right)\partial_y \delta u_y\,. 
\end{split}
\label{2.19}
\end{equation}
Similarly, we obtain the 
following fluctuations of the current
\begin{equation}
\begin{split}
&\delta J^t =\delta \rho= 
\left(\frac{\partial \rho}{\partial \mu}\right)_{T}
\delta \mu +
\left(\frac{\partial \rho}{\partial T}\right)_{\mu}
\delta T\,,\\
&\delta J^x =\rho \delta u^x\,,\qquad \delta J^y =\rho \delta u^y\,,\\
&\delta J^z= \rho \delta u^z+
\sigma_Q \left(-\partial_z \delta\mu+ \frac{\mu}{T}\partial_z \delta T\right)\,. 
\end{split}
\eqlabel{2.20}
\end{equation}
Substituting these expressions into equations of motion~\eqref{2.10}
and performing a Fourier transformation we get the following system of equations
\begin{equation}
\begin{split}
0=&\omega 
\left(\left(\frac{\partial \epsilon}{\partial \mu}\right)_{T}
\delta \mu+\left(\frac{\partial \epsilon}{\partial T}\right)_{\mu}\delta T\right)-
q (\e+P) \delta u^z\,,\\
0=&\omega (\epsilon+P)\delta u^z -q
\left( \left(\frac{\partial P}{\partial \mu}\right)_T \delta \mu
+\left(\frac{\partial P}{\partial T}\right)_{\mu}
\delta T\right)+i q^2(\frac 43 \eta +\zeta)\delta u^z\,,\\
0=&\omega 
\left(\left(\frac{\partial \rho}{\partial \mu}\right)_{T}\delta \mu+
\left(\frac{\partial \rho}{\partial T}\right)_{\mu}\delta T\right)
-q \rho \delta u^z + i q^2\sigma_Q \left(\delta \mu -\frac{\mu}{T}\delta T\right)\,,
\end{split}
\eqlabel{2.21a}
\end{equation}
\begin{equation}
\begin{split}
0=&\biggl(\omega (\epsilon+P)+i q^2 \eta\biggr) \delta u^x\,,\\
0=&\biggl(\omega (\epsilon+P)+i q^2 \eta\biggr) \delta u^y\,,\\
\end{split}
\eqlabel{2.21b}
\end{equation}
where we assembled equations in the decoupled sets. The three equations \eqref{2.21a} 
describe the propagation of sound waves in charged plasma, while the pair \eqref{2.21b}
describes the propagation of the shear modes of different polarizations.

In the shear channel, \eqref{2.21b}, the dispersion relation takes form 
\begin{equation}
\w\equiv i q^2 \cald=-i q^2\ \frac{\eta}{\e+P}=-i\frac {q^2}{T}\   
\frac{\eta}{s}\ \frac{T s}{ Ts -\mu \rho}  \,.
\eqlabel{sheard}
\end{equation}
Notice that even though in a holographic gauge theory plasma 
we study here the ratio of the shear viscosity to the entropy density 
is universal \cite{bbn}
\begin{equation}
\frac{\eta}{s}=\frac{1}{4\pi}\,,
\eqlabel{univer}
\end{equation}
a combination involving the diffusive constant $T \cald$ has a nontrivial 
dependence on $\frac T\mu$ and $\frac M\mu$. 

In the sound channel, \eqref{2.21a}, the dispersion relation takes form 
\begin{equation}
\begin{split}
\w=\pm c_s q -i \Gamma q^2+\calo(q^3)\,,
\end{split}
\eqlabel{sodisp}
\end{equation}
where the speed of sound $c_s$ and the attenuation $\Gamma$ are given by 
\begin{equation}
\begin{split}
c_s^2=&\biggl((\e+P)\ \frac{\del(P,\r)}{\del(T,\mu)}
+\r\ \frac{\del(\e,P)}{\del(T,\mu)}\biggr)\biggl((\e+P)\ \frac{\del(\e,\r)}
{\del(T,\mu)}\biggr)^{-1}\,,
\end{split}
\eqlabel{speed}
\end{equation}
\begin{equation}
\begin{split}
\Gamma=&\frac{2\eta}{3(\e+P)}\ \left(1+\frac{3\zeta}{4\eta}\right)
-\frac{\sigma_Q}{2T}\left(\frac{\del P}{\del \rho}\right)_\e
\left((\e+P)\ \frac{\del(P,\r)}{\del(T,\mu)}+\r \frac{\del (\e,P)}{\del(T,\mu)}
\right)^{-1}\times
\\
&\times \left((\e+P)\left(\left(\frac{\del \r}{\del \ln\mu}\right)_T
+\left(\frac{\del \r}{\del \ln T}\right)_\mu\right)-
\r\left(\left(\frac{\del \e}{\del \ln\mu}\right)_T
+\left(\frac{\del \e}{\del \ln T}\right)_\mu\right)
\right)\,.
\end{split}
\eqlabel{satt}
\end{equation}
It is instructive to analyze \eqref{sodisp} for conformal theories and for  theories 
with softly broken scale invariance, \ie, when 
\begin{equation}
\frac{M}{T}\ll 1\,,\qquad \frac{M}{\mu}\ll 1\,,
\eqlabel{soft}
\end{equation} 
where $M$ is a conformal symmetry breaking scale.
\nxt In a conformal hydrodynamics 
\begin{equation}
\zeta=0\,,\qquad \e=3P\,,\qquad \e\equiv T^4\ \cale\left(\frac{\mu}{T}\right)\,,\qquad 
 \r\equiv T^3\ \calr\left(\frac{\mu}{T}\right)\,,
\end{equation}
where $\cale$ and $\calr$ are functions of the dimensionless ratio $\frac \mu T$, thus 
\begin{equation}
\begin{split}
&\frac{\del(P,\rho)}{\del(T,\mu)}=\frac 13\ \frac{\del(\e,\rho)}{\del(T,\mu)}\,,\qquad 
\frac{\del(\e,P)}{\del(T,\mu)}=\frac 13\ \frac{\del(\e,\e)}{\del(T,\mu)}=0\,,\\
&\left(\frac{\del \r}{\del \ln\mu}\right)_T
+\left(\frac{\del \r}{\del \ln T}\right)_\mu=T^3\ \frac{\mu}{T}\ \calr'+3 T^3\ \calr+
T^4\ \left(-\frac {\mu}{T^2}\right)\ \calr' =3\r\,, \\
&\left(\frac{\del \e}{\del \ln\mu}\right)_T
+\left(\frac{\del \e}{\del \ln T}\right)_\mu=T^4\ \frac{\mu}{T}\ \cale'+4 T^4\ \cale+
T^5\ \left(-\frac {\mu}{T^2}\right)\ \cale' =4\e \,,\\
&\left(\frac{\del P}{\del \rho}\right)_\e=\frac{\del(P,\e)}{\del(\r,\e)}=\frac 13\ 
\frac{\del(\e,\e)}{\del(\r,\e)}=0\,.
\end{split}
\eqlabel{cftcdonst}
\end{equation}
As a result, we find 
for the speed of sound and the sound wave attenuation
\begin{equation}
c_s^2\bigg|_{CFT}=\frac 13\,,\qquad \Gamma\bigg|_{CFT}=\frac{2\eta}{3(\e+P)}
=\frac {2}{3T} \frac{\eta}{s}\ \frac{Ts}{Ts-\mu\r}\,.
\eqlabel{cfts}
\end{equation}
Once again, notice that $T \Gamma$ has a nontrivial dependence on $\frac{\mu}{T}$,
even though the shear viscosity ratio is universal \eqref{univer}\footnote{
In conformal holographic hydrodynamics this was first established in \cite{ss}.}.
\nxt When the scale invariance is softly broken \eqref{soft}, 
we find
\begin{equation}
\begin{split}
&c_s^2=\frac 13+\calo\left(\frac{M^2}{T^2}\right)\,,\\
&\Gamma=\frac{2\eta}{3(\e+P)}\left(1+\frac{3\zeta}{4\eta}\right)
+\frac{1}{T}\times \frac{\sigma_Q}{T}
\times \calo\left(\frac{M^2}{T^2}\right)\times  
\calo\left(\frac{M^2}{T^2}\right)\,,
\end{split}
\eqlabel{ncft}
\end{equation} 
where, given 
\eqref{cftcdonst}, we used 
\begin{equation}
\begin{split}
&\left(\frac{\del P}{\del \rho}\right)_\e=T\times 
\calo\left(\frac{M^2}{T^2}\right)\,,\\
&(\e+P)\left(\left(\frac{\del \r}{\del \ln\mu}\right)_T
+\left(\frac{\del \r}{\del \ln T}\right)_\mu\right)-
\r\left(\left(\frac{\del \e}{\del \ln\mu}\right)_T
+\left(\frac{\del \e}{\del \ln T}\right)_\mu
\right)\\
&\ \qquad =T^7\times \calo\left(\frac{M^2}{T^2}\right)\,.
\end{split}
\eqlabel{sau}
\end{equation}
In \eqref{ncft} and \eqref{sau} 
we suppressed dependence on $\frac {\mu}{T}$ in $\calo(M^2)$ terms. 
What is important for the later discussion is that the contribution 
of the term $\propto \sigma_Q$ to the attenuation $\Gamma$ 
is of order $\calo(M^4)$. 

\subsection{Sound waves of mass-deformed RN plasma}
We can use \eqref{speed} and \eqref{satt} to compute the 
speed of sound and the bulk viscosity of the mass-deformed 
RN plasma \eqref{ac5} to order $\calo(\l^2)$.

We find it convenient to introduce 
\begin{equation}
\ww\equiv \frac{\w}{2\pi T}\,,\qquad \qq\equiv \frac{q}{2\pi T}\,,
\eqlabel{dimwq}
\end{equation}
where $\{\w,q=|\vec q|\}$ are the frequency and the momentum of the sound 
mode. Furthermore, we parametrize the sound wave dispersion relation 
\eqref{sodisp} as 
\begin{equation}
\begin{split}
&\ww=\frac{\qq}{\sqrt{3}}\ \b_1
-i\frac{\qq^2}{3}\ \b_2+\calo(\qq^3)\,,\\
&\b_1\equiv  \sum_{n=0}^{1}\biggl\{\l^{2n}\ \b_{1,n}\biggr\}\,,\qquad \b_2=\sum_{n=0}^{1}\biggl\{\l^{2n}\ \b_{2,n}\biggr\}\,.
\end{split}
\eqlabel{sdisp}
\end{equation}
In this parametrization $\b_1=\b_2=1$ for a conformal plasma with vanishing chemical potential \cite{pss}.  
Using \eqref{ep}, \eqref{rhoa}, \eqref{ladef}, \eqref{univer} and introducing  
\begin{equation}
\begin{split}
&a_0=\frac{\b}{\sqrt{1+\k}}\,,\qquad a_1=-\b\sqrt{2}+\l^2 \b\ a_{1,2}(\k) \,,\\
&h_1=\frac{\k\sqrt{2}}{3\sqrt{1+\k}}+\l^2\ h_{1,2}(\k)\,,\qquad
g_0=\b\,,\qquad c_0=\l\,,\\
&c_{10}=c_{10}(\k)\,,
\end{split}
\eqlabel{uvk}
\end{equation}
for the UV parameters \eqref{uvparameters}, and 
\begin{equation}
\begin{split}
&a_2^h=\frac{\b}{(\k+2)\sqrt{1+\k}}+\l^2\b\ a_{2,2}^h(\k)\,,\qquad
g_0^h=\b+\l^2\b\ g_{0,2}^h(\k)\,,\\
&g_2^h=\l^2\b\  g_{2,2}^{h}(\k)\,,\qquad h_0^h=(1+\k)^{1/3}+\l^2\  h_{0,2}^{h}(\k)\,,\\
&c_0^h=\l\  c_{0,1}^{h}(\k)\,,
\end{split}
\eqlabel{irk}
\end{equation}
for the IR parameters \eqref{irparameters}, we find from \eqref{sodisp}-\eqref{satt}
\begin{equation}
\begin{split}
&\b_{1,0}=\pm {1}\,,\qquad \b_{2,0}=\frac{\k+2}{2\k+2}\,,\\
&\b_{1,1}=-\frac{\k\sqrt{2}}{144\sqrt{1+\k}}+\frac{1}{12}\ c_{10}(\k)\,,\\
\end{split}
\eqlabel{b01}
\end{equation}
while $\b_{2,1}$ directly determines the bulk viscosity to the shear viscosity ratio, 
since contribution to the attenuation proportional to the conductivity $\sigma_Q$
vanishes to order $\calo(\l^2)$, see \eqref{ncft},
and the conductivity itself is finite at criticality \cite{maeda1},
\begin{equation}
\begin{split}
&\frac{\zeta}{\eta}=\Delta(\k)\ \l^2+\calo(\l^4)=\Delta(\k)\ \frac{(\k+2)^2}{2^{3/2}(1+\k)^{3/2}\pi^2}\ \frac{M^2}{T^2}+
\calo\left(\frac{M^4}{T^4}\right)\,,
\end{split}
\eqlabel{ze}
\end{equation}
where
\begin{equation}
\begin{split}
&\Delta(\k)=\frac{ 8(1+\k)}{3(\k+2)}\ \beta_{2,1}-\frac{4 (1+\k)^{1/2} 2^{1/2}}{\k}\ h_{1,2}(\k)
+\frac 43 (\k+2) (1+\k)^{1/2}\ a_{2,2}^h(\k)\\
&-\frac{20}{3}\ g_{2,0}^h(\k)+\frac{4(\k+2)^2}{3(1+\k)}\ g_{2,2}^h(\k)
+\frac{2 (5 \k+4)}{(1+\k)^{1/3} \k}\ h_{0,2}^h(\k)
-\frac 16 (1+\k)^{1/3} \left(c_{0,1}^h(\k)\right)^2\,.
\end{split}
\eqlabel{defd}
\end{equation}

In the next section we study sound channel quasinormal modes  of mass-deformed RN black holes \eqref{back}
and compute $\b_1$ and $\b_2$ in \eqref{sdisp} to order $\calo(\l^2)$ inclusive, at $\k=2$. 
Since $c_{10}\bigg|_{\k=2}=c_{10}^{(0)}$ in Table \ref{tab1},  \eqref{b01} would provide a highly nontrivial test 
on a consistency of our thermodynamic and hydrodynamic analysis. Moreover, given that 
\begin{equation}
\begin{split}
&h_{1,2}\bigg|_{\k=2}=h_1^{(0)}\,,\qquad a_{2,2}^h\bigg|_{\k=2}=a_2^{h(0)}\,,\qquad  g_{2,0}^h\bigg|_{\k=2}=g_0^{h(0)}\,,\\
&g_{2,2}^h\bigg|_{\k=2}=g_2^{h(0)}\,,\qquad  h_{0,2}^h\bigg|_{\k=2}=h_0^{h(0)}\,,\qquad  c_{0,1}^h\bigg|_{\k=2}
=c_0^{h(0)}\,,
\end{split}
\eqlabel{reltotab}
\end{equation}
as presented in Table \ref{tab1} are finite, any possible divergence in the bulk viscosity at criticality,
\ie, at  $\k=\k_{c}$ \eqref{kc}, would arise from the divergence of $\b_{2,1}=\b_{2,1}(\k)$, as  ${\k\to 2}$.
The bulk viscosity to the shear viscosity ratio to leading order in $\frac{M^2}{T^2}$ at criticality 
is then computed from \eqref{ze}:
\begin{equation}
\lim_{T\to T_c}\ \frac{\zeta}{\eta}=\l^2\ \lim_{\k\to 2}\ \Delta(\k)+\calo(\l^4)=
\frac{2^{5/2}}{3^{3/2}\pi^2} \frac{M^2}{T_c^2}\ \lim_{\k\to 2}\Delta(\k)+\calo\left(\frac{M^4}{T_c^4}\right)\,.
\eqlabel{bulkc}
\end{equation}

\subsection{Fluctuations of the deformed RN black hole}\label{new}
To determine dispersion relation of the sound channel quasinormal mode we have to analyze 
fluctuation in the background geometry $\{g_{\mu\nu},A_\mu,\phi,\chi\}$  \eqref{back}:
\begin{equation}
\begin{split}
g_{\mu\nu}&\to g_{\mu\nu}+h_{\mu\nu}\,,\\
A_\mu&\to A_\mu+\delta A_\mu\,,\\
\phi&\to \phi+\delta \phi\,,\\
\chi&\to \chi+\delta\chi\,.
\end{split}
\eqlabel{deffl}
\end{equation}
We choose the gauge 
\begin{equation}
h_{tr}=h_{x_i r}=h_{rr}=0\,,\qquad \dd A_r=0\,.
\eqlabel{gauge}
\end{equation}
Additionally, we assume that all the fluctuations depend only on $(t,x_3,r)$, \ie, we have an $O(2)$
rotational symmetry in the $x_1x_2$ plane. At a linearized level, the fluctuations $\{h_{\mu\nu}, \dd A_\mu, \dd\phi, \dd \chi\}$
of different helicities with respect to this symmetry will decouple from each other. The sound channel 
quasinormal mode corresponds to helicity-zero fluctuations \cite{ks}:
\begin{equation}
\left\{h_{tt}\,, h_{aa}\equiv h_{xx}+h_{yy}\,, h_{tx_3}\,, h_{x_3x_3}\,, \dd A_t\,, \dd A_{x_3}\,, \dd \phi\,, 
\dd \chi \right\}\,.
\eqlabel{hel0}
\end{equation} 
We introduce
\begin{equation}
\begin{split}
h_{tt}=&e^{-i\w t+iq x_3}\ c_1^2\  H_{tt}\,,\\
h_{tx_3}=&e^{-i\w t+iq x_3}\ c_2^2\  H_{tz}\,,\\
h_{aa}=&e^{-i\w t+iq x_3}\ c_2^2\  H_{aa}\,,\\
h_{x_3x_3}=&e^{-i\w t+iq x_3}\ c_2^2\  H_{x_3x_3}\,,\\
\dd A_t=&e^{-i\w t+iq x_3}\ \cala_t\,,\\
\dd A_{x_3}=&e^{-i\w t+iq x_3}\ \cala_{x_3}\,,\\
\dd \phi=&e^{-i\w t+iq x_3}\ p\,,\\
\dd \chi=&e^{-i\w t+iq x_3}\ c\,,
\end{split}
\eqlabel{flf}
\end{equation}
where $\{H_{tt},H_{tx_3},H_{aa},H_{x_3x_3},\cala_{t},\cala_{x_3},p,c\}$ are functions of a radial coordinate  only. 
From the effective action \eqref{ac5} it is straightforward to derive 8 
second order differential equations of motion
for the fluctuations, and 4 first order differential constraints associated with fixing the gauge invariance 
as in \eqref{gauge}. Altogether we expect $8-4=4$ independent gauge-invariant combinations of fluctuations. Indeed, 
analyzing the transformations of \eqref{flf} under the residual gauge and diffeomorphism transformations  
it is straightforward to construct these combinations:
\begin{equation}
\begin{split}
Z_H=&4\frac{q}{\w} \ H_{tz}+2\ H_{zz}-H_{aa}\left(1-\frac{q^2}{\w^2}\frac{c_1'c_1}{c_2'c_2}\right)+2\frac{q^2}{\w^2}
\frac{c_1^2}{c_2^2}\ H_{tt}\,,\\
Z_\cala=&\cala_t+\frac{\w}{q}\ \cala_{x_3}-\frac{A'}{[\ln c_2^4]'}\ H_{aa}\,,\\
Z_p=&p-\frac{\phi'}{[\ln c_2^4]'}\ H_{aa}\,,\\
Z_c=&c-\frac{\chi'}{[\ln c_2^4]'}\ H_{aa}\,.
\end{split}
\eqlabel{physical}
\end{equation}
With somewhat tedious analysis we can verify that  equations of motion 
for the gauge-invariant fluctuations $\{Z_H,Z_\cala,Z_p,Z_c\}$ 
decouple.

The spectrum of quasinormal modes is determined \cite{ks} by imposing on $\{Z_H,Z_{\cala},Z_p,$ $Z_c\}$ 
an incoming wave boundary condition at the horizon, and requiring vanishing of the non-normalizable modes for 
$\{Z_H,Z_{\cala},Z_p,Z_c\}$
near the boundary. In the hydrodynamic limit $\ww\to 0$, $\qq\to 0$ with $\frac \ww\qq$ kept fixed, 
this leads to the following 
perturbative expansions
\begin{equation}
\begin{split}
Z_H=&(1-x)^{-i\ww}\biggl(z_{H,0}+i\qq\ z_{H,1}+\calo(\qq^2)\biggr)\,,\\
Z_\cala=&(1-x)^{-i\ww}\biggl(z_{\cala,0}+i\qq\ z_{\cala,1}+\calo(\qq^2)\biggr)\,,\\
Z_c=&(1-x)^{-i\ww}\biggl(z_{p,0}+i\qq\ z_{p,1}+\calo(\qq^2)\biggr)\,,\\
Z_c=&(1-x)^{-i\ww}\biggl(z_{c,0}+i\qq\ z_{c,1}+\calo(\qq^2)\biggr)\,,
\end{split}
\eqlabel{expan}
\end{equation}
with the following boundary conditions on $\{z_{H,i},z_{\a,i},z_{\c,i}\}$:    
\begin{equation}
\begin{split}
&\lim_{x\to 1_-} z_{H,0}=1\,,\qquad \lim_{x\to 1_-}z_{H,1}=0\,,\qquad \lim_{x\to 1_-} z_{\cala,i}=\lim_{x\to 1_-} z_{p,i}
=\lim_{x\to 1_-} z_{c,i}={\rm finite}\,,\\
&z_{H,i}=\calo(x)\,,\qquad z_{\cala,i}= \calo\left(x^{1/2}\right)\,,\qquad z_{p,i}=\calo\left(x^{1/2}\right)\,,
\qquad z_{c,i}=\calo\left(x^{3/4}\right)\,,
\end{split}
\eqlabel{bc}
\end{equation}
as $x\to 0_+$.

To leading order in the hydrodynamic approximation, wave functions of the gauge-invariant 
fluctuations $\{z_{H,0},z_{\cala,0},z_{p,0},z_{c,0}\}$ 
satisfy the following equations
\begin{equation}
\begin{split}
0=&z_{H,0}''+\calc_{101}\ z_{H,0}'+\calc_{102}\ z_{\cala,0}'+\calc_{103}\ z_{p,0}'+\calc_{104}\ z_{c,0}'
+\calc_{105}\ z_{H,0}+\calc_{106}\ z_{\cala,0}\\
&+\calc_{107}\ z_{p,0}+\calc_{108}\ z_{c,0}\,,\\
0=&z_{\cala,0}''+\calc_{201}\ z_{H,0}'+\calc_{202}\ z_{\cala,0}'+\calc_{203}\ z_{p,0}'+\calc_{204}\ z_{c,0}'
+\calc_{205}\ z_{H,0}+\calc_{206}\ z_{\cala,0}\\
&+\calc_{207}\ z_{p,0}+\calc_{208}\ z_{c,0}\,,\\
0=&z_{p,0}''+\calc_{301}\ z_{H,0}'+\calc_{302}\ z_{\cala,0}'+\calc_{303}\ z_{p,0}'+\calc_{304}\ z_{c,0}'
+\calc_{305}\ z_{H,0}+\calc_{306}\ z_{\cala,0}\\
&+\calc_{307}\ z_{p,0}+\calc_{308}\ z_{c,0}\,,\\
0=&z_{c,0}''+\calc_{401}\ z_{H,0}'+\calc_{402}\ z_{\cala,0}'+\calc_{403}\ z_{p,0}'+\calc_{404}\ z_{c,0}'
+\calc_{405}\ z_{H,0}+\calc_{406}\ z_{\cala,0}\\
&+\calc_{407}\ z_{p,0}+\calc_{408}\ z_{c,0}\,,
\end{split}
\eqlabel{order0}
\end{equation}
where connection coefficients $\calc_{i0j}$ are complicated nonlinear functionals of the background 
fields $\{g,A,\phi,\chi\}$ with explicit dependence on $x$
and $\b_{12}\equiv \b_1^2$, and implicit dependence on $\k$ and $\l$:
\begin{equation}
\calc_{i0j}=\calc_{i0j}\biggl[\{g,A,\phi,\chi\};\ x;\ \b_{12}\biggr]\,.
\eqlabel{calc0}
\end{equation}
To leading order in the hydrodynamic approximation, wave functions of the gauge-invariant 
fluctuations $\{z_{H,1},z_{\cala,1},z_{p,1},z_{c,1}\}$ 
satisfy  equations identical to \eqref{order0}, apart from the source terms $\{\calj_H,\calj_{\cala},\calj_p,\calj_c\}$:
\begin{equation}
\begin{split}
0=&z_{H,1}''+\calc_{101}\ z_{H,1}'+\calc_{102}\ z_{\cala,1}'+\calc_{103}\ z_{p,1}'+\calc_{104}\ z_{c,1}'
+\calc_{105}\ z_{H,1}+\calc_{106}\ z_{\cala,1}\\
&+\calc_{107}\ z_{p,1}+\calc_{108}\ z_{c,1}+\calj_H\,,\\
0=&z_{\cala,1}''+\calc_{201}\ z_{H,1}'+\calc_{202}\ z_{\cala,1}'+\calc_{203}\ z_{p,1}'+\calc_{204}\ z_{c,1}'
+\calc_{205}\ z_{H,1}+\calc_{206}\ z_{\cala,1}\\
&+\calc_{207}\ z_{p,1}+\calc_{208}\ z_{c,1}+\calj_{\cala}\,,\\
0=&z_{p,1}''+\calc_{301}\ z_{H,1}'+\calc_{302}\ z_{\cala,1}'+\calc_{303}\ z_{p,1}'+\calc_{304}\ z_{c,1}'
+\calc_{305}\ z_{H,1}+\calc_{306}\ z_{\cala,1}\\
&+\calc_{307}\ z_{p,1}+\calc_{308}\ z_{c,1}+\calj_p\,,\\
0=&z_{c,1}''+\calc_{401}\ z_{H,1}'+\calc_{402}\ z_{\cala,1}'+\calc_{403}\ z_{p,1}'+\calc_{404}\ z_{c,1}'
+\calc_{405}\ z_{H,1}+\calc_{406}\ z_{\cala,1}\\
&+\calc_{407}\ z_{p,1}+\calc_{408}\ z_{c,1}+\calj_c\,,
\end{split}
\eqlabel{order1}
\end{equation}
with
\begin{equation}
\begin{split}
\calj_H=&\calc_{111}\ z_{H,0}'+\calc_{112}\ z_{\cala,0}'+\calc_{113}\ z_{p,0}'+\calc_{114}\ z_{c,0}'
+\calc_{115}\ z_{H,0}+\calc_{116}\ z_{\cala,0}\\
&+\calc_{117}\ z_{p,0}+\calc_{118}\ z_{c,0}\,,\\
\calj_\cala=&\calc_{211}\ z_{H,0}'+\calc_{212}\ z_{\cala,0}'+\calc_{213}\ z_{p,0}'+\calc_{214}\ z_{c,0}'
+\calc_{215}\ z_{H,0}+\calc_{216}\ z_{\cala,0}\\
&+\calc_{217}\ z_{p,0}+\calc_{218}\ z_{c,0}\,,\\
\calj_p=&\calc_{311}\ z_{H,0}'+\calc_{312}\ z_{\cala,0}'+\calc_{313}\ z_{p,0}'+\calc_{314}\ z_{c,0}'
+\calc_{315}\ z_{H,0}+\calc_{316}\ z_{\cala,0}\\
&+\calc_{317}\ z_{p,0}+\calc_{318}\ z_{c,0}\,,\\
\calj_c=&\calc_{411}\ z_{H,0}'+\calc_{412}\ z_{\cala,0}'+\calc_{413}\ z_{p,0}'+\calc_{414}\ z_{c,0}'
+\calc_{415}\ z_{H,0}+\calc_{416}\ z_{\cala,0}\\
&+\calc_{417}\ z_{p,0}+\calc_{418}\ z_{c,0}\,.
\end{split}
\eqlabel{sources}
\end{equation}
The new connection coefficients $\calc_{i1j}$ are complicated nonlinear functionals of the background 
fields $\{g,A,\phi,\chi\}$ with explicit dependence on $x$
and $\b_{12},\b_2$, and implicit dependence on $\k$ and $\l$:
\begin{equation}
\calc_{i1j}=\calc_{i1j}\biggl[\{g,A,\phi,\chi\};\ x;\ \{\b_{12}\,,\b_2\}\biggr]\,.
\eqlabel{calc1}
\end{equation}
Explicit expressions for $\calc_{ikj}$ are available from the author upon the request.

It is straightforward to determine the asymptotic expansions (satisfying \eqref{bc}) and thus set up the boundary value 
problem what would determine $\{\b_{12},\b_2\}$, along with the coefficients of the normalizable modes for 
 $\{z_{H,i},z_{\cala,i},z_{p,i},z_{c,i}\}$. We found that, unfortunately, directly solving the resulting boundary value problem 
(as it was done for example  in \cite{bpa}) is not possible with reasonable computational resources.
In the rest of this section we explain the origin of the problem and
outline the solution\footnote{A simple explicit example explaining the use of a new computational technique is 
presented in Appendix \ref{sbh}.}. 

Recall that \eqref{order0} and \eqref{order1} were obtained in the hydrodynamic limit from the full
quasinormal mode equations for $\{Z_H,Z_\cala,Z_p,Z_c\}$. The latter equations have a structure identical to 
that of \eqref{order0}. For example, we have
\begin{equation}
\begin{split}
0=&Z_{H}''+\calc_{H,1}\ Z_{H}'+\calc_{H,2}\ Z_{\cala}'+\calc_{H,3}\ Z_{p}'+\calc_{H,4}\ Z_{c}'
+\calc_{H,5}\ Z_{H}+\calc_{H,6}\ Z_{\cala}\\
&+\calc_{H,7}\ Z_{p}+\calc_{H,8}\ Z_{c}\,,\\
\end{split}
\eqlabel{Zh}
\end{equation}
with  
\begin{equation}
\calc_{H,i}=\calc_{H,i}\biggl[\{g,A,\phi,\chi\};\ x;\ \{\ww,\qq\}\biggr]\,.
\end{equation}
Explicit expressions for $\calc_{H,i}$ show that (some of) these coefficients have a simple pole at  
\begin{equation}
0=\ww^2-(1-x)^2 \qq^2\,.
\eqlabel{polepos}
\end{equation}
The residues of these poles are always proportional to the derivative of the background gauge 
potential\footnote{This explains why such poles are absent in the analysis of \cite{bls} or \cite{bca}.} 
$A'$, or $(A')^2$, (see \eqref{back}), and does not vanish when either (or both) $\phi=1$ and $\chi=0$. 
In other words, these poles {\it always} occur in studies of the sound channel quasinormal modes of charged black
holes in asymptotic $AdS_5$ geometry.  Since the speed of sound squared $c_s^2$
 in  holographic plasma with a UV conformal fixed 
point is bounded by \cite{csb1,csb2}\footnote{See \cite{bp2} for an exception.} 
\begin{equation}
c_s^2\le \frac 13\,,
\eqlabel{csb}
\end{equation}
from \eqref{polepos} we always expect to encounter a singularity in the connection coefficients 
of the differential equations describing the propagation of the sound channel quasinormal modes 
of the charged black holes inside the range of the integration, \ie, for $x\in (0,1)$. Of course, 
the singularity in the coefficients of the differential equation does not imply that the solutions 
are singular --- in our case we find that they are not --- however, the presence of such singularities 
poses  technical difficulties  for the boundary value problem one has to solve. Actually, the issue is even more 
complicated: the successive hydrodynamic approximations produce equations (\eqref{order0} and \eqref{order1}
 in our case) which connection coefficients involve successive derivatives with respect to $c_s$. 
Thus, while some of the coefficients $\calc_{i0j}$ in \eqref{order0} have a simple pole inside 
the range of integration,  some of the coefficients $\calc_{i1j}$ in \eqref{sources} have a double pole!
Finally, a perturbative expansion in the deformation parameter $\l$, which is needed to extract $\b_{2,1}$
and ultimately the bulk viscosity of the charged plasma, see \eqref{bulkc}, produces yet higher order poles --- 
here we need to deal with the third order poles in the connection coefficients inside the range of 
the integration in the boundary value problem. We find that numerical techniques for solving the boundary value problem 
developed in \cite{abk} become unreliable  once the connection coefficients of the differential equations 
have a second or higher order poles. A new approach is needed.
   
Basically, we need to reformulate the boundary value problem in such a way that connection coefficients of the 
corresponding differential equations have simple poles inside the range of integration, at worst. 
First, even though we  are after the hydrodynamics to order $\calo(\l^2)$ only, we treat \eqref{order0}
and \eqref{order1} exactly in $\l$. The boundary value problem would then determine
\begin{equation}
\b_1=\b_1(\l)\,,\qquad \b_2=\b_2(\l)\,.
\eqlabel{b1b2e}
\end{equation} 
From the data sets for small $\l$ we can extract $\b_{1,0},\b_{1,1},\b_{2,0},\b_{2,1}$ (see \eqref{sdisp}).
Such a step insures that $\calc_{i0j}$ have simple poles at most, while $\calc_{i1j}$ have double poles at most.
In particular, the boundary value problem for \eqref{order0} is amenable to the treatment of \cite{abk}.
Second, the hydrodynamic origin of the second order poles in $\calc_{i1j}$ implies that 
the residues of the second order poles in $\calc_{i1j}$ and $\frac{\del \calc_{i0j}}{\del \b_{12}}$ 
must be proportional to each other. Indeed, for each value of indexes $\{i,j\}$ we find
\begin{equation}
\calc_{i1j}=-2\b_2 \frac{\sqrt{\b_{12}}}{\sqrt{3}}\ \frac{\del}{\del \b_{12}} \calc_{i0j}+\tc_{i1j}\,,\qquad 
\tc_{i1j}=\tc_{i1j}\biggl[\{g,A,\phi,\chi\};\ x;\ \b_{12}\biggr]\,,
\eqlabel{c1c0rel}
\end{equation}
with $\tc_{i1j}$ having at most simple poles for $x\in (0,1)$. Given \eqref{c1c0rel}, the boundary value problem 
\eqref{order1} can be reformulated as follows. 
\nxt We represent 
\begin{equation}
\begin{split}
z_{H,1}=&-2\b_2 \frac{\sqrt{\b_{12}}}{\sqrt{3}}\ \frac{\del}{\del\b_{12}} \hz_{H,0}+\tz_{H,1}\,,\\
z_{\cala,1}=&-2\b_2 \frac{\sqrt{\b_{12}}}{\sqrt{3}}\ \frac{\del}{\del\b_{12}} \hz_{\cala,0}+\tz_{\cala,1}\,,\\
z_{p,1}=&-2\b_2 \frac{\sqrt{\b_{12}}}{\sqrt{3}}\ \frac{\del}{\del\b_{12}} \hz_{p,0}+\tz_{p,1}\,,\\
z_{c,1}=&-2\b_2 \frac{\sqrt{\b_{12}}}{\sqrt{3}}\ \frac{\del}{\del\b_{12}} \hz_{c,0}+\tz_{c,1}\,.
\end{split}
\eqlabel{breakz}
\end{equation}
\nxt The wave functions $\{\hz_{H,0},\hz_{\cala,0},\hz_{p,0},\hz_{c,0}\}$ satisfy exactly the same equations 
as $\{z_{H,0},$ $z_{\cala,0},z_{p,0},z_{c,0}\}$, \ie, \eqref{order0}, with $\b_{12}$ being treated as an extra free parameter
and the only change in the boundary conditions being 
\begin{equation}
\lim_{x\to 0_+} \hz_{H,0}={\rm finite}\equiv \calz_0(\b_{12})\,.
\eqlabel{dervalue}
\end{equation}
Clearly, imposing the Dirichlet boundary condition  $\calz_0(\b_{12})=0$ would determine 
the value of $\b_{12}$ which would identify all $\hz_{\cdots,0}$ with $z_{\cdots,0}$:
\begin{equation}
\hz_{\cdots,0}\bigg|_{\calz_0(\b_{12})=0}=z_{\cdots,0}\,,
\eqlabel{ident}
\end{equation}
where $\cdots$ stand for any of $\{H,\cala,p,c\}$. For each value of $\l$, we 
can evaluate $\calz_0$ for some set of $\b_{12}$ and compute 
\begin{equation}
\calz_0'\equiv \frac{\del}{\del \b_{12}} \calz_0(\b_{12})\bigg|_{\calz_0(\b_{12})=0}\,.
\eqlabel{derz0}
\end{equation}
The boundary value problem(s) implementing this procedure involve solving ODE's with 
simple poles at most in the connection coefficients inside the range of integration.
\nxt Given \eqref{breakz} and the definition of $\hz_{\cdots,0}$, it is straightforward to see that 
$\tz_{\cdots,1}$ would satisfy equations identical to \eqref{order1}, except that now the source terms 
\eqref{sources} would be constructed from the connection coefficients $\tc_{i1j}$. For example,
\begin{equation}
\begin{split}
\calj_{H}\to \tilde{\calj}_{H}\ \equiv\  &\tc_{111}\ z_{H,0}'+\tc_{112}\ z_{\cala,0}'+\tc_{113}\ z_{p,0}'+\tc_{114}\ z_{c,0}'
+\tc_{115}\ z_{H,0}+\tc_{116}\ z_{\cala,0}
\\
&+\tc_{117}\ z_{p,0}+\tc_{118}\ z_{c,0}\,.
\end{split}
\eqlabel{defjh}
\end{equation}
By construction, this final boundary value problem would involve ODE's with simple poles at most in the 
connection coefficients inside the range of integration. However, since it does not depend on 
$\b_2$ (see \eqref{c1c0rel}), 
to find a solution we must modify\footnote{This is the only modification in the boundary conditions.}
the Dirichlet condition at the boundary for $\tz_{H,1}$:
\begin{equation}
\lim_{x\to 0_+} \tz_{H,1}={\rm finite}\equiv \calz_1\,.
\eqlabel{dervalue1}
\end{equation}
\nxt Ultimately, $z_{H,1}$ must satisfy the Dirichlet condition at the boundary \eqref{bc}. 
Using \eqref{breakz}, \eqref{dervalue}, 
\eqref{derz0} and \eqref{dervalue1} this leads to 
\begin{equation}
-2\b_2 \frac{\sqrt{\b_{12}}}{\sqrt{3}}\ \calz_0'+\calz_1=0\,,
\eqlabel{condb2}
\end{equation}
which determines $\b_2$ as
\begin{equation}
\b_2=\frac{\sqrt{3}}{2\sqrt{\b_{12}}}\ \frac{\calz_1}{\calz_0'}\,.
\eqlabel{b2def}
\end{equation}
Thus, we succeeded in reformulating the boundary value problem for computing $\b_2$
in such a way that it does not involve numerical integration of ODE's with connection 
coefficients having singularities stronger than simple poles for $x\in (0,1)$.

We emphasize once again that the  procedure described above is necessary only in computing  sound 
channel quasinormal modes for charged black holes in asymptotic $AdS_5$ geometry. 
On the other hand, it is generic, and can be applied to AdS-Schwarzschild black holes 
as well. In Appendix \ref{sbh} we demonstrate the new method for the computation of the dispersion relation of the sound waves 
in strongly coupled $\caln=4$ SYM plasma.

\subsection{Mass-deformed RN plasma transport at criticality}

In this section we present results for the speed of sound and the bulk viscosity of the 
mass-deformed RN plasma in the vicinity of the second order phase transition. 
We perform numerical analysis for a fixed $\k=2$ as a function of (a small) mass-deformation 
parameter $\l$, see \eqref{ladef}. In general (see \eqref{kc})
\begin{equation}
\k_c=\k_c(\l)=2+\calo(\l^2)\,.
\eqlabel{kcc}
\end{equation} 
However, setting $\k=2$ is still sufficient to extract the leading correction to the speed of sound
\begin{equation}
\lim_{T\to T_c}\ \left(c_s^2\right)+\calo(\l^4)=\frac{1}{3}\ \left(\b_{1}(\k=2,\l)\right)^2\,,
\eqlabel{cstc}
\end{equation}
and the leading contribution 
to the bulk viscosity near the vicinity of the phase transition, see \eqref{bulkc}:
\begin{equation}
\begin{split}
\lim_{T\to T_c}\ \left(\l^{-2}\ \times\ \frac{\zeta}{\eta}\right)+\calo(\l^2)=\lim_{\l\to 0} \ \Delta(\k=2,\l)\,.
\end{split}
\eqlabel{bulkcc}
\end{equation}

\subsubsection{The speed of sound}

\begin{figure}[t]
\begin{center}
\psfrag{l}{{$\l$}}
\psfrag{b12}{{$(1-\b_{1}^2)$}}
  \includegraphics[width=4in]{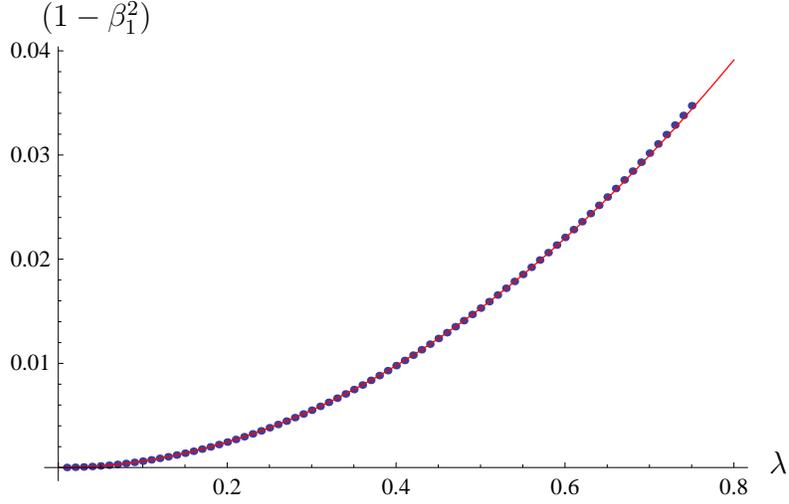}
\end{center}
  \caption{(Colour online)
Deviation of the speed of sound $(1-\b_1^2)$, see \eqref{sdisp},  
in mass-deformed RN plasma from its conformal value as a function of the mass-deformation parameter 
$\l$ at $\k=2$, see \eqref{ladef}. The blue dots represents results obtained from the 
holographic hydrodynamic equations \eqref{order0}, and the solid red line represents 
thermodynamic prediction, see \eqref{b01}. 
} \label{figure1}
\end{figure}

First, 
we  verify that in the conformal limit 
the speed of sound is independent of the chemical potential, see \eqref{cfts} . We find 
\begin{equation}
\bigg|\left(\b_{1}(\k,\l=0)\right)^2-1\bigg|\ \sim 10^{-14}\cdots 10^{-11}\,,\qquad \k\in[0.4,2]\,.
\eqlabel{cftres}
\end{equation} 

Figure \ref{figure1} presents the results for the deviation of the speed of sound in RN plasma from the conformal value 
as a function of the mass-deformation parameter $\l$: the blue dots are obtained from directly solving the holographic 
hydrodynamic equations to leading order \eqref{order0}, 
 while the solid red line\footnote{We used \eqref{b01} and the result of Table \ref{tab1}: $c_{10}(\k=2)=c_{10}^{(0)}$.}
\begin{equation}
(1-\b_{1}^2)\bigg|_{red}\equiv -2\b_{1,1}\ \l^2\ =\ 0.061132(8)\,,
\eqlabel{defred}
\end{equation}
 is the thermodynamic prediction for this deviation, 
valid in the limit $\l\to 0$. 
The results are in excellent agreement: for instance, for 
$\l=0.01$ we find 
\begin{equation}
\bigg|\frac{(1-\b_{1}^2)_{red}}{(1-\b_{1}^2)_{blue}}-1\bigg|\ \approx\ 5\times 10^{-6}\,.
\eqlabel{accub12}
\end{equation}

\subsubsection{The bulk viscosity}

\begin{figure}[t]
\begin{center}
\psfrag{cz0}{{$\calz_0$}}
\psfrag{db12}{{$\left(b_{12}-b_{12}^\star\right)$}}
  \includegraphics[width=4in]{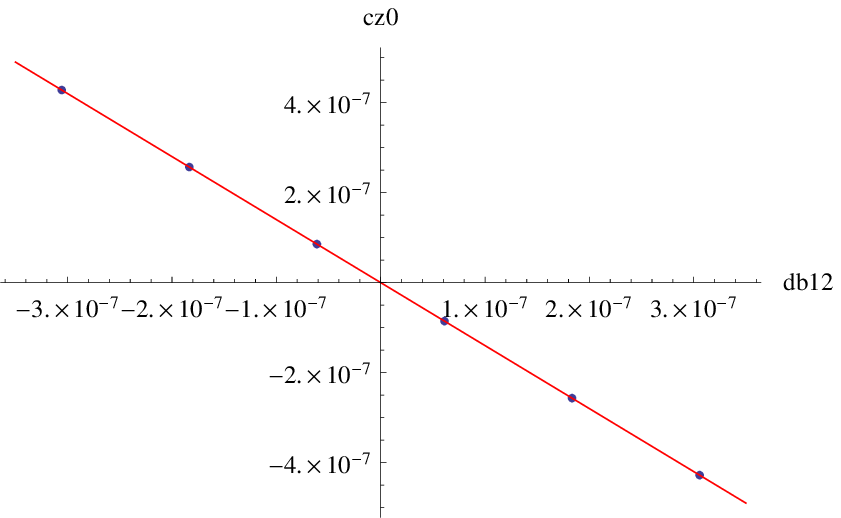}
\end{center}
  \caption{(Colour online)
Blue dots represent $\calz_0(\b_{12})$ (see \eqref{dervalue}) for the mass-deformation parameter $\l=0.01$. 
$\b_{12}^\star\equiv\left(\b_1(\l=0.01)\right)^2$. The solid red line represents the best linear fit to $\calz_0$.} \label{figure2}
\end{figure}

\begin{figure}[t]
\begin{center}
\psfrag{la}{{$\l$}}
\psfrag{beta2}{{$\b_2$}}
  \includegraphics[width=4in]{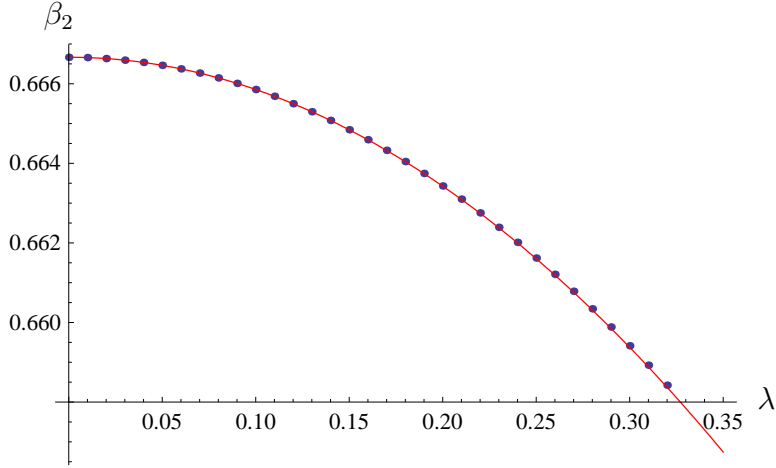}
\end{center}
  \caption{(Colour online)
Blue dots represent $\b_2(\l)$ (see \eqref{b2def}) at $\k=2$. 
The solid red line represents the best quadratic fit to the first 10 blue dots. } \label{figure3}
\end{figure}

Scale invariance of the model at $\l=0$ predicts \eqref{b01} 
\begin{equation}
\b_2(\l=0)_{prediction}=\b_{2,0}(\k=2)=\frac 23 \,.
\eqlabel{pred}
\end{equation} 
Explicit computation of the attenuation coefficient $\b_2$ following the method explained in
section \ref{new}  yields
\begin{equation}
\bigg|\frac{\b_2(\l=0)}{\b_2(\l=0)_{prediction}}-1\bigg|\ \approx\  2\times 10^{-6}\,.
\eqlabel{bulkacc}
\end{equation}

Before we present a general plot for the bulk viscosity in RN plasma, we discuss in some details results for 
$\l=0.01$. Blue dots in Figure \ref{figure2}  represent the results for $\calz_0(\b_{12})$ 
(see \eqref{dervalue}) for values of $\b_{12}$ in the vicinity of $\b_{12}^\star\equiv \left(\b_{1}(\l=0.01)\right)^2$ 
as evaluated from 
solving the hydrodynamic equations at leading order \eqref{order0}. We 
present the data as a function of  
$(\b_{12}-\b_{12}^\star)$ since, following the discussion around \eqref{dervalue}, we expect 
\begin{equation}
\calz_0(\b_{12}=\b_{12}^\star)=0\,.
\eqlabel{b12star}
\end{equation} 
The solid red line represents the best linear fit to the hydrodynamic $\calz_0$ data. Explicitly, we find
\begin{equation}
\calz_0\bigg|_{red}=-8.\times 10^{-14} - 1.400011(9)\ \left(\b_{12}-\b_{12}^\star\right)\,.
\eqlabel{fitz0}
\end{equation}
From \eqref{fitz0} we determine
\begin{equation}
\calz_0'\bigg|_{\l=0.01}=- 1.400011903986774\,.
\eqlabel{z0p001}
\end{equation}
Next, we evaluate $\calz_1$ as explained in section \ref{new}:
\begin{equation}
\calz_1\bigg|_{\l=0.01}=-1.0777132662397266\,,
\eqlabel{z1l001}
\end{equation}
which following \eqref{b2def} computes $\b_2$:
\begin{equation}
\b_2\bigg|_{\l=0.01}=0.6666585596689777\,.
\eqlabel{b2sing}
\end{equation} 

Figure \ref{figure3} presents results (blue points) for $\b_2(\l)$, were we simply iterated the procedure 
described above for $\b_2({\l=0.01})$. The solid red line represents the best quadratic fit (with the linear term absent)
to the first 10 blue points:
\begin{equation}
\b_2\bigg|_{red}=0.666666(3) - 0.081002(5)\ \l^2\,.
\eqlabel{b2fit}
\end{equation} 
Note that $\b_2(\l=0)$ in this fit is remarkably close to exact CFT value \eqref{pred}.
We can now compute  $\b_{2,1}$ in \eqref{sdisp} as 
\begin{equation}
\b_{2,1}=\frac 12 \frac{d^2}{d\l^2}\ \b_2\bigg|_{red}\qquad \Longrightarrow\ \qquad \b_{2,1}=- 0.081002(5)\,.
\eqlabel{b21gen}
\end{equation}
Finally, using results in Table \ref{tab1}, we find from \eqref{defd}
\begin{equation}
\Delta=0.224133(3) + 2 \b_{2,1}\,,
\eqlabel{compdd}
\end{equation} 
thus, from \eqref{ze}
\begin{equation}
\frac{\zeta}\eta\ \approx\ 0.062128(2)\ \l^2+\calo(\l^4)\,.
\eqlabel{bulksing}
\end{equation}
Curiously, since 
\begin{equation}
\frac 13-c_s^2\equiv  \frac 13 (-2\b_{1,1})=\frac 13 (1-\b_{12}^\star)= 0.020377(6)\ \l^2+\calo(\l^4)\,,
\eqlabel{speedbound}
\end{equation}
the bulk viscosity bound for holographic gauge theory plasma proposed in \cite{bbulk}
\begin{equation}
\frac{\zeta}{\eta}\ge 2\left(\frac 13-c_s^2\right)\,,
\eqlabel{boundbulk}
\end{equation}
is satisfied in the vicinity of the phase transition of mass-deformed RN plasma.

To summarize, we find:
\begin{equation}
\frac{\zeta}{\eta}=3.0488(5)\ \left(\frac 13-c_s^2\right)+\calo\left(\left(\frac 13-c_s^2\right)^2\right)\,,
\eqlabel{result}
\end{equation}
for the bulk viscosity of the mass-deformed RN plasma in the vicinity of the second order phase transition. 

\section{Dynamical critical phenomena in RN plasma}
In this section we study the response of conformal and mass-deformed RN plasma to inhomogeneous 
and time-dependent variation of the chemical potential
\begin{equation}
\mu\to \mu+\dd\mu(t,\vec{x})\,,\qquad 
\dd\mu(t,\vec{x})=\int\frac{d^3k}{(2\pi)^3}\int\ \frac{d\w}{2\pi}\ e^{i\vec{k}\cdot \vec{x}-i\w t}\ 
\mu_{\w,\vec{k}}\,.
\eqlabel{chempot}
\end{equation}
At a linearized level the variation of the chemical potential would produce a corresponding 
variation in the charge density, $\dd\r(t,\vec{x})$ ( $\r_{\w,\vec{k}}$ for the Fourier components). As in 
conventional theory of dynamical critical 
phenomena  \cite{hh}, we introduce the dynamical susceptibility $\c_{\w,\vec{k}}$,
\begin{equation}
\c_{\w,\vec{k}}=\left(\frac{ \r_{\w,\vec{k}}}{ \mu_{\w,\vec{k}}}\right)\bigg|_{T}\,,\qquad \lim_{(\w,\vec{k})\to 0}\ 
\c_{\w,\vec{k}}=\c_T=\left(\frac{\del\r}{\del \mu}\right)\bigg|_T\,.
\eqlabel{chidyn}
\end{equation}
The fluctuation-dissipation theorem states that 
\begin{equation}
G(\w,\vec{k})=\frac{2T}{\w}\ \Im \c_{\w,\vec{k}}\,,
\eqlabel{fdt}
\end{equation}
where $G(\w,\vec{k})$ is a Fourier transform of the charge density variation two-point 
correlation function 
\begin{equation}
G(t,\vec{x})=\langle\dd\r(t,\vec{x})\dd\r(0,\vec{0})\rangle_{\mu=0}\,.
\eqlabel{2pt}
\end{equation} 
Furthermore, the equal-time correlation function 
\begin{equation}
G(\vec{k})\equiv G(\w=0,\vec{k})\,,
\eqlabel{2pte}
\end{equation} 
is related to the static susceptibility 
\begin{equation}
\c_{\vec{k}}\equiv \c_{\w=0,\vec{k}}\,,
\eqlabel{2ptes}
\end{equation} 
by the equipartition theorem
\begin{equation}
G(\vec{k})=T \c_{\vec{k}}\,.
\eqlabel{ept}
\end{equation} 

In the vicinity of (but not at) the critical point $t=\frac{T}{T_c}-1\to 0$  ($t\ne 0$)
  the equal-time correlation function 
$G(\vec{x})$ decays exponentially 
\begin{equation}
G(\vec{x})\propto e^{-|\vec{x}|/\xi}\,,
\eqlabel{decay}
\end{equation}
where $\xi$ is the correlation length, implying that $G(\vec{k})$, and  through the  equipartition 
relation \eqref{ept} the static susceptibility $\c_{\vec{k}}$, have a pole at 
\begin{equation}
k^2\propto -\xi^{-2}\,.
\eqlabel{stp}
\end{equation} 
Right at the critical point, $t=0$, the equal-time correlation function has a power-law decay 
\begin{equation}
G(\vec{x})\propto |\vec{x}|^{-p+2-\eta}\qquad \Rightarrow\qquad G(\vec{k})\propto |\vec{k}|^{-2+\eta}\,,
\eqlabel{power}
\end{equation}
where $p=3$ is the number of spatial dimensions and $\eta$ is the anomalous scaling exponent.

The theory of dynamical critical phenomena \cite{hh} predicts that 
in the vicinity of the continuous phase transition and for $|\vec{k}|\sim \xi^{-1}$ 
the full dynamical susceptibility $\c_{\w,\vec{k}}$
will develop a pole at  
\begin{equation}
\w \propto -i \xi^{-z}\,,
\eqlabel{dynpole}
\end{equation}
with $z$ being the dynamical critical exponent of the system. The frequency
in \eqref{dynpole} defines a relaxation time $\tau^{-1}$ 
\begin{equation}
\t^{-1}\equiv i\w\propto \xi^{-z}\,,
\eqlabel{deftau}
\end{equation}
characterizing the equilibration time scale of the dynamical system.

In the rest of this section we analyze dynamical susceptibility of the strongly coupled 
conformal RN plasma. Following up the poles in static susceptibility (see \eqref{stp}) 
in the vicinity of the phase transition we determine the (static) scaling exponent $\nu$
of the correlation length: $\xi\propto t^{-\nu}$. The scaling of the pole 
in the dynamical susceptibility (see \eqref{dynpole}) determines the dynamical critical exponent 
of the RN plasma. Finally, we comment on dynamical critical phenomena in 
mass-deformed RN plasma.

\subsection{Computation of $\c_{\w,\vec{k}}$ in holographic dual}
Without the loss of generality we can assume that 
\begin{equation}
k^i=q\ \dd^i_3\,.
\eqlabel{defkv}
\end{equation}
In dual gravitational description the variation of the chemical potential $\mu_{\w,k}$ translates into the 
variation of the non-normalizable mode of the fluctuation of the bulk vector field $\cala_t$ \eqref{flf}, 
correspondingly the variation of the non-normalizable mode in the 
gauge-invariant fluctuation $Z_\cala$ \eqref{physical}. 
Up to an overall factor\footnote{This factor is finite in the vicinity of the transition.} 
(see \eqref{scalesuv}), the  variation in the charge density $\r_{\w,k}$ is  the 
normalizable component of $\cala_t$ (or $Z_\cala$ for the gauge-invariant fluctuation). 
The fluctuations of the dual gravitational background were analyzed extensively in section 
\ref{fluctuations}. As we study here the dynamical critical phenomena of the conformal RN plasma, 
we can consistently set $Z_c=0$. We are left with the linear coupled system of fluctuations 
$\{Z_H,Z_\cala,Z_p\}$ \eqref{physical}.  
There are two important differences in analysis of this system of fluctuations compare to the one 
in section \ref{fluctuations}, relevant for the computation of the sound wave dispersion relation:
\nxt In both cases  $\{Z_H,Z_\cala,Z_p\}$ must satisfy an incoming boundary conditions 
at the horizon. In case of the sound waves, we had  to impose the vanishing of all the 
non-normalizable modes for  $\{Z_H,Z_\cala,Z_p\}$ at the boundary (see \eqref{bc}).
As a result, solving the boundary value problem determined the dispersion relation for the 
sound waves \eqref{sdisp}
\begin{equation}
\ww=\ww(\qq)\,.
\eqlabel{souwd}
\end{equation} 
In computing the dynamical susceptibility, both  $\ww$ and $\qq$ are independent; nonetheless, 
the boundary value problem on the gauge-invariant fluctuations has a solution because the non-normalizable 
component of $Z_\cala$ at the boundary is nonzero now.    
\nxt Transport coefficients are encoded in the perturbative in $\qq$ expansion of the sound wave dispersion 
relation \eqref{sodisp}. Thus, it was sufficient to implement the perturbative 
(hydrodynamic) expansion for the fluctuations \eqref{expan} and the dispersion relation \eqref{sdisp}.
In study of the dynamical critical phenomena of strongly coupled RN plasma we are interested in 
the poles (for complex $\ww\,, \qq$) of the dynamical susceptibility $\c_{\ww,\qq}$. Thus, 
we can not do the computations perturbatively in $\ww\,, \qq$. Indeed, since 
the system of equations of motion for the fluctuations $\{Z_H,Z_\cala,Z_p\}$ is linear, without   
the loss of generality we can set the non-normalizable component of $Z_\cala$ near the boundary, 
namely $\mu_{\ww,\qq}$
to one:
\begin{equation}
\mu_{\ww,\qq}=1\,.
\eqlabel{muto1}
\end{equation}
From \eqref{chidyn} we have 
\begin{equation}
\c_{\ww,\qq}=\r_{\ww\,\qq}\,,
\eqlabel{ssus}
\end{equation}
thus any poles in the  dynamical susceptibility must come from the 
poles in the normalizable component of $Z_\cala$ near the boundary, and as such 
they must be non-perturbative 
in $\ww\,, \qq$.

We can summarize now the boundary value problem whose solution would determine the dynamical susceptibility.
Introducing\footnote{The $\ww-$ and $\qq-$dependent rescaling are for convenience in further analysis.} 
\begin{equation}
\begin{split}
Z_H=&(1-x)^{-i\ww}\ \ww^{-2}\ z_h(x,\ww,\qq)\,,\\
Z_\cala=&(1-x)^{-i\ww}\ \qq^{-2}\ z_\cala(x,\ww,\qq)\,,\\
Z_p=&(1-x)^{-i\ww}\ \qq^{-2}\ z_p(x,\ww,\qq)\,,
\end{split}
\eqlabel{expsus}
\end{equation}
the equations of motion for $\{z_h,z_\cala,z_p\}$ are solved with the following boundary conditions:
\begin{equation}
\begin{split}
&\lim_{x\to 1_-}z_H=\lim_{x\to 1_-}z_\cala=\lim_{x\to 1_-}z_p={\rm finite}\,,\\
&z_H=\calo(x)\,,\qquad z_\cala=1+\calz(\ww,\qq)\ x^{1/2}+\calo(x)\,,\qquad z_p=\calo(x^{1/2})\,,\qquad {\rm as}\ 
x\to 0_+\,.
\end{split}
\eqlabel{ssbc}
\end{equation}
The normalizable component of $z_\cala$ near the boundary $\calz$ is proportional to the dynamical 
susceptibility:
\begin{equation}
\c_{\ww,\qq}\propto \calz(\ww,\qq) \,.
\eqlabel{czrel}
\end{equation}

\subsection{Static and dynamical susceptibilities of RN plasma}

\begin{figure}[t]
\begin{center}
\psfrag{k}{{$\k$}}
\psfrag{chi}{{$\calz^{-1}\propto \c_{\ww=0,\qq=0}^{-1}$}}
  \includegraphics[width=4in]{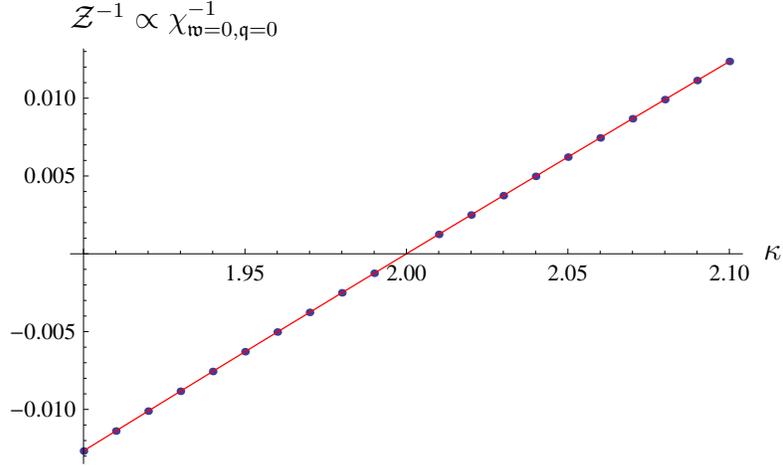}
\end{center}
  \caption{(Colour online) The scaling (blue dots) of the inverse of the static susceptibility 
$\c_{\ww=0,\qq=0}$ in the vicinity
of the critical point. The solid red line is a quadratic fit to the data. 
} \label{figure4}
\end{figure}

\begin{figure}[t]
\begin{center}
\psfrag{k}{{$\k$}}
\psfrag{q2}{{$\qq_*^2$}}
  \includegraphics[width=4in]{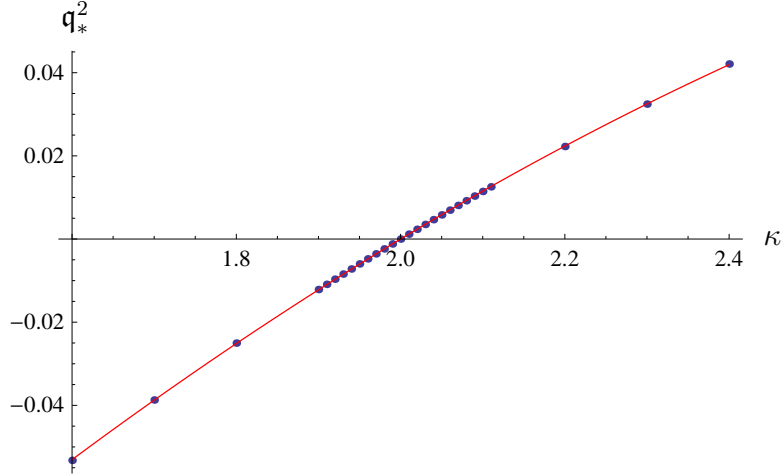}
\end{center}
  \caption{(Colour online) Poles of the static susceptibility in the vicinity of the 
critical point: $\c_{\ww=0,\qq=\qq_*}^{-1}=0$. The solid red line is a quadratic fit to the data.
} \label{figure5}
\end{figure}

\begin{figure}[t]
\begin{center}
\psfrag{k}{{$\qq^2$}}
\psfrag{chi}{{$(\calz^{crit})^{-1}\propto (\c_{c,\ww=0,\qq}^{crit})^{-1}$}}
  \includegraphics[width=4in]{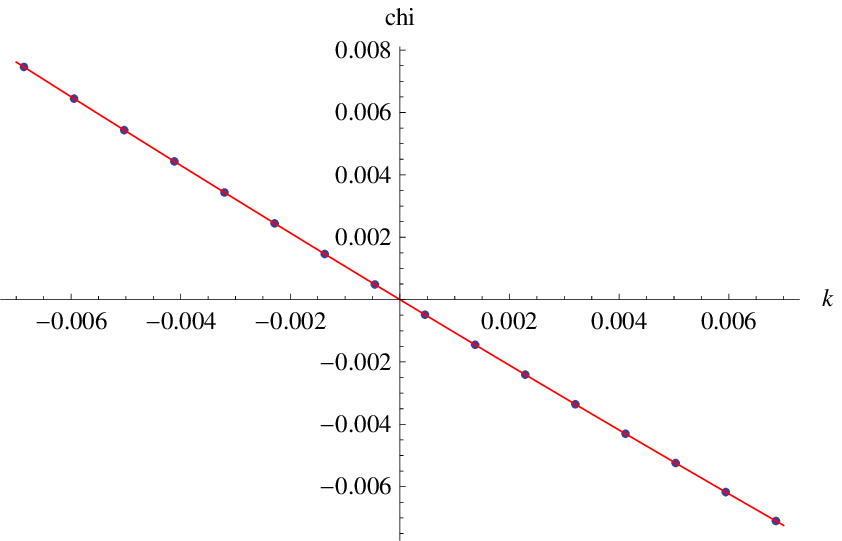}
\end{center}
  \caption{(Colour online) The scaling (blue dots) of the inverse of the static susceptibility 
$\c_{\ww=0,\qq}^{crit}$ at  the critical point, $\k=2$. The solid red line is a quadratic (in $\qq^2$) fit to the data. 
} \label{figure6}
\end{figure}

\begin{figure}[t]
\begin{center}
\psfrag{k}{{$\k$}}
\psfrag{tau}{{$i\ \frac{\ww_*}{\qq^2}$}}
  \includegraphics[width=4in]{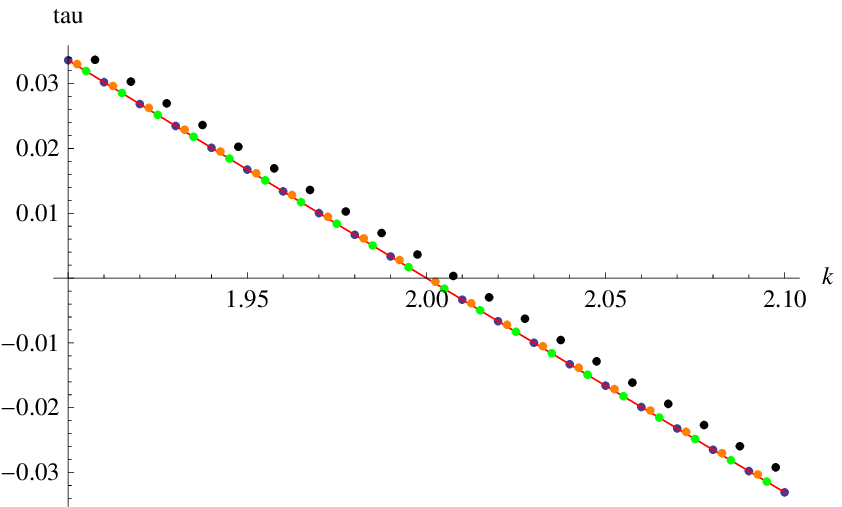}
\end{center}
  \caption{(Colour online) Poles of the dynamical susceptibility in the vicinity of the critical point, 
$\c_{\ww=\ww_*,\qq}^{-1}=0$ for a set of momenta values $\qq^2=$\ : $10^{-6}$ (blue dots) , $10^{-5}$ (green dots),
$10^{-4}$ (orange dots) and $10^{-3}$ (black dots). The solid red line is a quadratic fit to $i \frac{\ww_*}{\qq^2}$ 
at $\qq^2=10^{-6}$.
} \label{figure7}
\end{figure}

As discussed in section \ref{therm0}, a second order phase transition in strongly coupled RN plasma 
happens for 
\begin{equation}
\k=2\qquad \Longleftrightarrow\qquad  \frac{T}{\mu}=\frac{\sqrt{2}}{\pi}\,.
\end{equation}
We do not provide the technical details of the analysis of the boundary value problem 
\eqref{expsus}-\eqref{czrel} and present only the results. 
We point out  that the coefficients of the differential equations for 
$\{z_h,z_\cala,z_p\}$ are non-singular as $\k\to 2$; thus the appearance of poles in $\calz$
in this limit is not obvious.  
 
Figure \ref{figure4} shows the inverse of the static susceptibility at $\qq=0$ (blue dots) in the vicinity
of the critical  point. The solid red line represents the best quadratic fit to the data:
\begin{equation}
\calz^{-1}_{fit}=4.74744\ \cdot\ 10^{-8}+0.125116\ (\k-2)-0.0156647\ (\k-2)^2+\calo((\k-2)^3)\,.
\eqlabel{red4}
\end{equation}
 The red line \eqref{red4} 
intersects the $\k$ axis at 
\begin{equation}
\k_c=1.999999(6)\,,
\eqlabel{kcsus}
\end{equation}
in excellent agreement with the expected value $\k_c=2$.
Thus we reproduce the thermodynamic result for the static susceptibility
\begin{equation}
\c_{\ww=0,\qq=0}=\chi_T\ \propto\ \calz\ \propto\  \frac{1}{\k-\k_c}\ \propto\ +t^{-1/2}\,,
\qquad |\k-\k_c|\ll \k_c\,,
\eqlabel{thermorepr}
\end{equation}
where we used relation \eqref{kappacrit} between $\k$ and the reduced temperature $t$.

Figure \ref{figure5} presents the poles (blue dots) of the static susceptibility at $\qq=\qq_*$
in the vicinity of the critical point:
\begin{equation}
\c_{\ww=0,\qq=\qq_*}^{-1}=0\,.
\eqlabel{statpoles}
\end{equation}
The solid red line represents the best quadratic fit to the data:
\begin{equation}
q_{*,fit}^2=-2.32509\ \cdot\ 10^{-6} + 0.11873\ (\k-2) - 0.0347648\ (\k-2)^2+\calo((\k-2)^2)\,.
\eqlabel{fitstatic}
\end{equation}
Notice that in the stable phase, \ie, for  $\k\le 2$, in the vicinity of the 
phase transition the poles in the static susceptibility are for purely imaginary momenta,
which implies the exponential decay of the charge density two-point correlation function
\eqref{decay}. Furthermore, from \eqref{stp} we identify the correlation length 
as 
\begin{equation}
(2\pi T_c\ \xi)^2\ \propto\ \qq_*^{-2}\ \propto\ \frac{1}{\k-\k_c}\ \propto +t^{-1/2}\,,\qquad 
0<\k_c-\k\ll \k_c\,,
\eqlabel{xiscale}
\end{equation} 
where we used the results of the fit \eqref{fitstatic} and the relation between $\k$ 
and the reduced temperature $t$ \eqref{kappacrit}. From \eqref{xiscale} we extract the 
(static) critical exponent $\nu$: 
\begin{equation}
\xi\ \propto\ t^{-\nu}\ \propto\ t^{-1/4}\qquad \Rightarrow\qquad \nu=\frac 14 \,.
\eqlabel{xiscale2}
\end{equation}
Given that the static critical exponent $\a=\frac 12$, \eqref{xiscale2} implies 
that the hyperscaling relation is violated
\begin{equation}
2-\a\ \ne p\ \nu\,,
\eqlabel{hyperscale}
\end{equation}
where $p=3$ stands for the number of spatial dimensions of the system.

Figure \ref{figure6} shows the inverse of the static susceptibility as a function of $\qq$ (blue dots) 
right at the critical point $\k=2$. The solid red line represents the best quadratic (in $\qq^2$) fit to the data 
\begin{equation}
(\calz^{crit}_{fit})^{-1}=-1.57468\ \cdot 10^{-8} - 1.06109\ \qq^2 + 3.84182\ \qq^4+\calo(\qq^6)\,.
\eqlabel{red6}
\end{equation}
 The red line \eqref{red6} 
intersects the $\qq^2$ axis at 
\begin{equation}
\qq^2_c=-1.57468\ \cdot 10^{-8}\,,
\eqlabel{qcsus}
\end{equation}
in excellent agreement with the expected value $\qq_c^2=0$ \eqref{power}. 
The data implies  
\begin{equation}
\c^{crit}_{\ww=0,\qq}\ \propto\ \calz^{crit}\ \propto \qq^{-2}\qquad \Longleftrightarrow\qquad
  \c^{crit}_{\ww=0,\qq}\ \propto\ 
\qq^{-2+\eta}\,,
\eqlabel{defeta}
\end{equation}
which determines the anomalous critical exponent $\eta$ as 
\begin{equation}
\eta=0\,.
\eqlabel{etares}
\end{equation}

Figure \ref{figure7} presents the poles in the dynamical susceptibility at $\ww=\ww_*$ in the vicinity of the critical point 
\begin{equation}
\c^{-1}_{\ww=\ww_*,\qq}=0\,,
\eqlabel{polesdyn} 
\end{equation}
for select values of the momenta $\qq$:
\begin{equation}
\qq^2=\{10^{-6}\,, 10^{-5}\,, 10^{-4}\,, 10^{-3}\}\qquad \sim\qquad \{{\rm blue}\,, {\rm green}\,, {\rm orange}\,, 
{\rm black}\}\,.
\eqlabel{select}
\end{equation}
The results of the analysis clearly show that as $\qq\to 0$ the values of $i \frac{\ww_*}{\qq^2}$ tend to a universal profile 
\begin{equation}
\lim_{\qq\to 0}\ i\ \frac{\ww_*}{\qq^2}=2.79163\ \cdot 10^{-6} - 0.333392 (\k-2) + 0.0278087 (\k-2)^2+\calo((\k-2)^3)\,,
\eqlabel{reddyn}
\end{equation}
which is presented by the solid red line on Figure \ref{figure7}\footnote{In practice we used the best quadratic 
fit to to $i \frac{\ww_*}{\qq^2}$ at $\qq^2=10^{-6}$. }. Given \eqref{reddyn} we can determine the critical scaling 
of the relaxation time (see \eqref{deftau}) of strongly coupled RN plasma
\begin{equation}
(2\pi T_c\ \t)^{-1}\equiv i\ww_*\ \propto\ \qq^2\cdot (\k-\k_c)\ \propto\ (2\pi T_c\ \qq\xi)^{2}\cdot (2\pi T_c\ \xi)^{-4}\ \propto\ 
(2\pi T_c\ \xi)^{-4} \,,
\eqlabel{taufin}
\end{equation} 
where we wrote the $\qq$ dependence as $\propto \qq \xi$ and  used \eqref{xiscale}. Thus,
\begin{equation}
\t\ \propto\ \xi^z\ \propto\ \xi^4\qquad \Rightarrow\qquad z=4\,.
\eqlabel{zres}
\end{equation}
In \cite{maeda1} Maeda, Natsuume and Okamura argued that the strongly coupled RN plasma at criticality should be identified 
with the 'model B' according to classification of \cite{hh}. As such, the dynamical critical exponent $z$ in this model is predicted to be 
\cite{hh}
\begin{equation}
z=4-\eta\,.
\eqlabel{zpred}
\end{equation} 
Since $\eta=0$ \eqref{etares} for the strongly coupled RN plasma, we explicitly confirm the conclusion of \cite{maeda1}.

\subsection{Universality class of the mass-deformed RN plasma}

\begin{figure}[t]
\begin{center}
\psfrag{k}{{$\tilde{\k}$}}
\psfrag{q2}{{$\qq_*^2$}}
  \includegraphics[width=4in]{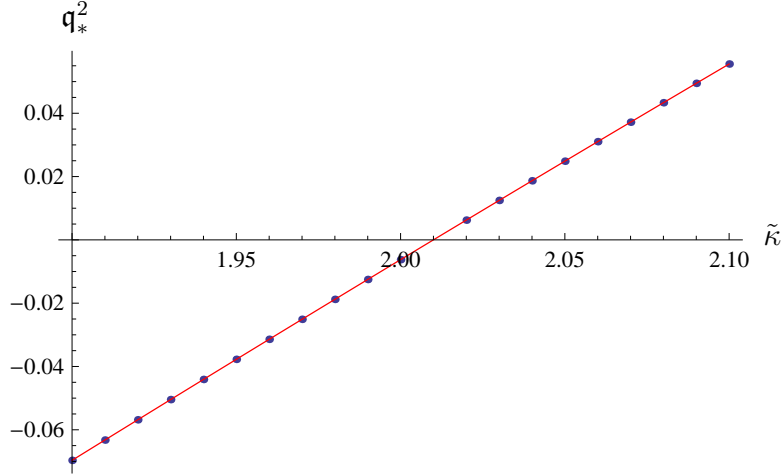}
\end{center}
  \caption{(Colour online) Poles of the static susceptibility of mass-deformed RN plasma 
in the vicinity of the critical point: $\c_{\ww=0,\qq=\qq_*}^{-1}=0$. The solid red line is a quadratic fit to the data.
} \label{figure8}
\end{figure}

\begin{figure}[t]
\begin{center}
\psfrag{k}{{$\qq^2$}}
\psfrag{chi}{{$(\calz^{crit})^{-1}\propto (\c_{c,\ww=0,\qq}^{crit})^{-1}$}}
  \includegraphics[width=4in]{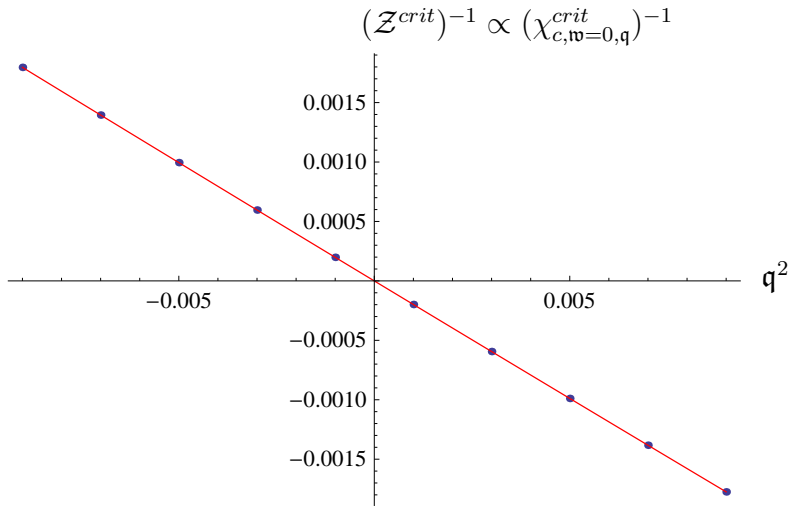}
\end{center}
  \caption{(Colour online) The scaling (blue dots) of the inverse of the static susceptibility 
of mass deformed RN plasma $\c_{\ww=0,\qq}^{crit}$ at  the critical point, 
$\tilde{\k}=\tilde{\k}_c$,
\eqref{tkc}. The solid red line is a quadratic (in $\qq^2$) fit to the data. 
} \label{figure9}
\end{figure}

\begin{figure}[t]
\begin{center}
\psfrag{k}{{$\tilde{\k}$}}
\psfrag{tau}{{$i\ \frac{\ww_*}{\qq^2}$}}
  \includegraphics[width=4in]{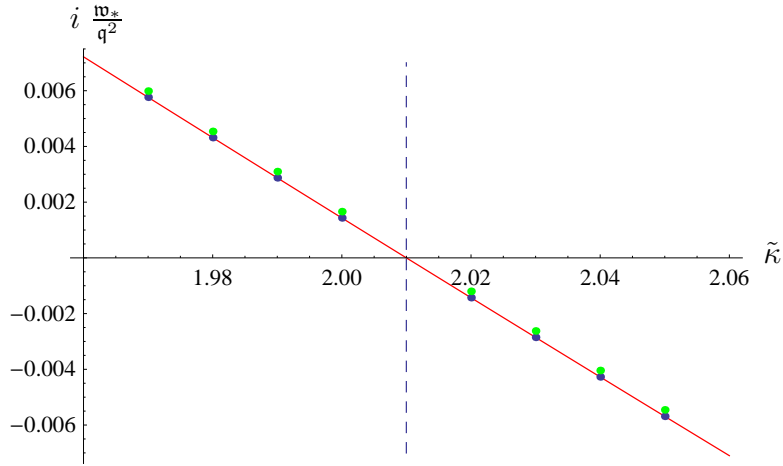}
\end{center}
  \caption{(Colour online) Poles of the dynamical
 susceptibility of mass deformed 
RN plasma in the vicinity of the critical point, 
$\c_{\ww=\ww_*,\qq}^{-1}=0$ for a set of momenta values $\qq^2=$\ : $10^{-6}$ (the solid red line 
quadratic fit) , $10^{-5}$ (blue dots),
and $10^{-3}$ (green dots). The vertical dashed blue line denotes $\tilde{\k}=\tilde{\k}_c$, 
see \eqref{tkc}. 
} \label{figure10}
\end{figure}

In section 2.5 we computed the static critical exponents $(\a,\b,\gamma, \delta)$
of the conformal RN plasma, see \eqref{statcritn4}.   
Further, in previous section we determined the  remaining static critical exponents 
$(\nu,\eta)$ of this theory, and determined its dynamical critical exponent $z$.

Notice that the mass deformation of the $\caln=4$ SYM plasma modifies its infrared 
(hydrodynamic) properties --- it generates a non-vanishing bulk viscosity.
Thus, one might worry that it is a relevant deformation at criticality 
and potentially might change the universality class of the theory.   
We show here that this is not the case: the universality class of the mass 
deformed $\caln=4$ SYM is the same as that of the conformal theory. 

We already argued that some of the static critical exponents of the mass deformed theory 
are unchanged (see section 2.6):
\begin{equation}
\left(\a\,,\b\,,\gamma\,,\delta\right)=\left(\frac 12\,,\frac 12\,,\frac 12\,, 2\right)\,.
\eqlabel{stmd}
\end{equation} 
To determine the remaining critical exponents $(\nu,\eta,z)$ we have to generalize the 
computation of the dynamical susceptibility as explained in section 4.1 to include
the fluctuation  $Z_c$ in addition to $\{Z_H,Z_\cala,Z_p\}$.  In analogy to 
\eqref{expsus}  we introduce  
\begin{equation}
Z_c=(1-x)^{-i \ww}\ \qq^{-2}\ z_c(x,\ww,\qq) \,,
\eqlabel{expzc}
\end{equation}
and solve equations of motion for $\{z_h,z_\cala,z_p,z_c\}$ with boundary conditions
\eqref{ssbc} supplemented with 
\begin{equation}
\lim_{x\to 1_-} z_c={\rm finite}\,,\qquad {\rm and}\qquad  
z_c=\calo(x^{3/4})\,,\qquad {\rm as}\ x\to 0_+\,.
\eqlabel{ssbczc}
\end{equation}
As before, up to an overall constant, the dynamical susceptibility is identified 
with the normalizable mode of $z_\cala$ near the asymptotic $AdS_5$ boundary, see
\eqref{czrel}.  We present the results of such analysis for one specific choice of
$\frac{M}{T_c}$, namely, 
\begin{equation}
\frac{M}{T_c}\approx 10^{-1}\times \frac{\pi\sqrt{3}}{2}=0.272(1)\,.
\eqlabel{mtcnum}
\end{equation}
We use approximate sign since relation \eqref{ladef}, which was used 
to obtain \eqref{mtcnum}, receives 
corrections of order $\frac{M^2}{T_c^2}$.

Figure \ref{figure8} is equivalent to Figure \ref{figure5} for the conformal RN 
plasma. Clearly, the slope of the solid red line (representing the quadratic fit to the 
data) is finite at $\qq_*^2=0$, which, much like in \eqref{fitstatic}-\eqref{xiscale2},
implies that the critical exponent $\nu=\frac 14$.  Note a technical detail:
$\tilde{\k}=\k+\calo(M^2/T_c^2)$; the precise relation is not important for the 
purpose of extracting the critical exponent. From the intersection of the solid 
red line the $\tilde{\k}$ axis we find 
\begin{equation}
\tilde{\k}_c =2.0099887(0)\,.
\eqlabel{tkc}
\end{equation}

Figure \ref{figure9} is equivalent to Figure \ref{figure6} for the conformal RN 
plasma. It represents the scaling of the inverse static susceptibility of the mass-deformed 
RN plasma at criticality. As in \eqref{defeta}, we conclude that $\eta=0$ in this case.

Finally, Figure \ref{figure10} is equivalent to Figure \ref{figure7} for the conformal RN 
plasma. As in \eqref{reddyn}-\eqref{zres} we conclude that $z=4$ for the dynamical critical
exponent of mass-deformed RN plasma.

\section{Conclusions}

In this paper we presented extensive analysis of the critical phenomena in 
superconformal $\caln=4$ SYM gauge theory plasma at finite temperature 
and a single $U(1)\subset SU(4)$ R-symmetry chemical potential. 
From the explicit analysis of the dynamical susceptibility near criticality 
we extracted the static critical exponents $(\nu,\eta)$ and identified 
the static universality class of the model\footnote{This corrects 
conclusions of \cite{stu2}.}:
\begin{equation}
\left(\a\,, \b\,, \gamma\,, \delta\,, \nu\,, \eta\right)=\left(\frac 12\,, 
\frac 12\,, \frac 12\,,
2\,, \frac 14\,, 0\right)\,.
\eqlabel{finalstat}
\end{equation} 
We explicitly computed the dynamical critical exponent of the theory 
\begin{equation}
z=4\,,
\eqlabel{finalz}
\end{equation}
and confirmed the identification of the $\caln=4$ SYM plasma dynamical
 universality 
class with that of 'model B' according to classification of \cite{hh}, originally made in
\cite{maeda1}. 

We demonstrated that although a deformation of the $\caln=4$ SYM theory by a dimension-3 
operator is relevant in the infrared --- in particular, it generates the non-zero 
bulk viscosity in the effective hydrodynamic description of the mass-deformed 
plasma ---  the static and the dynamical universality classes of the theory remains 
unchanged: \eqref{finalstat} and \eqref{finalz}.

We carefully studied the propagation of the sound waves in non-conformal  charged plasma.
We confirmed the computation of the speed of sound from the 
thermodynamic analysis with the direct result extracted from the quasinormal mode 
analysis of the holographic dual. We showed that 
the bulk viscosity of the mass-deformed $\caln=4$ SYM plasma remains 
finite at criticality,  and satisfies the bulk viscosity bound \cite{bbulk} 
(at least to order $\calo\left(M^2/T_c^2\right)$). Our computations challenge 
the Onuki's model  \cite{bulk3}  for the behavior of bulk viscosity near criticality.
Indeed, the latter model predicts
\begin{equation}
\zeta\bigg|_{Onuki}\propto |t|^{-z\nu+\a} \propto |t|^{-1/2} \,,
\eqlabel{onpred}
\end{equation}
for the mass-deformed $\caln=4$ SYM plasma, in contradiction with the finite 
result we obtained. Thus, when combined 
with analysis in \cite{bp3}, it appears that as of now, there is no model of 
transport at continuous phase transitions that is not in conflict with direct 
(first principle) holographic computations. It is important to use 
gauge theory/string theory correspondence to develop a consistent model. 

In order to explain holographic computations of transport at criticality 
it might be necessary to generalize the framework of near-equilibrium relaxation.
Specifically, it appears necessary to formulate the theory of dynamical critical 
phenomena in which different non-equilibrium correlators relax to equilibrium with different 
dynamical critical exponents.

\section*{Acknowledgments}
I would like to thank Ofer Aharony, Micha Berkooz and Rob Myers for valuable discussions.
I further thank Rob Myers for comments on the 
manuscript. I would  like to thank Mitchel Institute for Fundamental 
Physics and Astronomy and Weizmann Institute for hospitality 
during the various stage of this project.
Research at Perimeter Institute is
supported by the Government of Canada through Industry Canada and by
the Province of Ontario through the Ministry of Research \&
Innovation. I gratefully acknowledge further support by an NSERC
Discovery grant and support through the Early Researcher Award
program by the Province of Ontario. 

\appendix
\section{Coefficients $\calc_{ij}$}\label{appa}

\begin{equation}
\begin{split}
&\calc_{11}=\frac{1}{(x-1) (H_0^6-1)^2 (2+H_0^3) (1+\k)}\ (H_0^{15} (1+\k)-H_0^{12} (1+\k)
-2 H_0^9\\
&\times
 (3 \k^2 (x-1)^2
+\k+1)-2 H_0^6 (5 \k^2 (x-1)^2
-\k-1)
-H_0^3 (8 \k^2 (x-1)^2-\k-1)\\
&-1-\k)\,.
\end{split}
\eqlabel{c11}
\end{equation}
\begin{equation}
\begin{split}
\calc_{12}=&\frac{3(2 H_0^3+1) \b}{2(x-1) (H_0^3-1) (2+H_0^3) H_0}\,.
\end{split}
\eqlabel{c12}
\end{equation}
\begin{equation}
\begin{split}
\calc_{13}=&\frac{(2 H_0^3+1) H_0^3 \k}{\sqrt{1+\k} (x-1) (H_0^3-1) (2+H_0^3)}\,.
\end{split}
\eqlabel{c13}
\end{equation}
\begin{equation}
\begin{split}
\calc_{14}=&\frac{\b}{4(2+H_0^3)}\,.
\end{split}
\eqlabel{c14}
\end{equation}
\begin{equation}
\begin{split}
\calc_{15}=&-\frac{2 (2 H_0^3+1)) \k^2 H_0^3}{(2+H_0^3) (H_0^6-1) (H_0^3-1) (1+\k)}\,.
\end{split}
\eqlabel{c15}
\end{equation}
\begin{equation}
\begin{split}
\calc_{16}=&\frac{2 H_0^2 (5 H_0^6-2) \b (2 H_0^3+1) \k^2}{(2+H_0^3) (H_0^6-1)^2 (H_0^3-1) (1+\k)}\,.
\end{split}
\eqlabel{c16}
\end{equation}
\begin{equation}
\begin{split}
&\calc_{17}=\frac{(2 H_0^3+1)(1+\k+H_0^3((x-1)^2\k^2-2-2\k)+H_0^6(1+\k)) \k^2 H_0^7 \b m^2}{12(1+\k)^2 (H_0^6-1)^2 (H_0^3-1)^2 (2+H_0^3)}\,.
\end{split}
\eqlabel{c17}
\end{equation}
\begin{equation}
\begin{split}
&\calc_{21}=\frac{2H_0}{3\b (H_0^6-1) (2+H_0^3) (x-1) (1+\k)} (3 H_0^9 (1+\k)+H_0^6 (8 \k^2 (x-1)^2-3-3 \k)\\
&+H_0^3 
(4 \k^2 (x-1)^2-3-3 \k)+3+3 \k)\,.
\end{split}
\eqlabel{c21}
\end{equation}
\begin{equation}
\begin{split}
&\calc_{22}=-\frac{ 1}
{3(2+H_0^3) (H_0^6-1)^2 (x-1) (1+\k)}(3 H_0^{15} (1+\k)+H_0^{12} (8 \k^2 (x-1)^2-3 \k-3)
\\
&+2 H_0^9 (11 \k^2 (x-1)^2-3-3\k)+2 H_0^6 (11 \k^2 (x-1)^2+3 \k+3)
+H_0^3 (3 \k+3\\
&+20 \k^2 (x-1)^2)-3-3 \k)\,.
\end{split}
\eqlabel{c22}
\end{equation}
\begin{equation}
\begin{split}
&\calc_{23}=\frac{2 H_0^4 \k }{3\b (1+\k)^{3/2} (2+H_0^3) (H_0^6-1) (H_0^3+1) (x-1)}(3 H_0^9 (1+\k)-H_0^6 (2 \k^2
 (x-1)^2\\
&-3-3 \k)-H_0^3 (3+3 \k+4 \k^2 (x-1)^2)-3-3 \k)\,.
\end{split}
\eqlabel{c23}
\end{equation}
\begin{equation}
\begin{split}
\calc_{24}=&-\frac{H_0 (H_0^3-1)}{6(2+H_0^3)}\,.
\end{split}
\eqlabel{c24}
\end{equation}
\begin{equation}
\begin{split}
&\calc_{25}=-\frac{4 H_0^4 \k^2 }{3(2+H_0^3) (H_0^6-1)^2 (H_0^3+1) \b (1+\k)^2}(3 H_0^9 (1+\k)-H_0^6 (2 \k^2 (x-1)^2\\
&-3-3 \k)-H_0^3 (3+3 \k+4 \k^2 (x-1)^2)
-3-3 \k)\,.
\end{split}
\eqlabel{c25}
\end{equation}
\begin{equation}
\begin{split}
&\calc_{26}=-\frac{4 H_0^3 \k^2}{9(1+\k)^2 (H_0^6-1)^3 (H_0^3-1) (2+H_0^3)} (18+18 \k+2 H_0^3 (8 \k^2 (x-1)^2-9 \k-9)
\\
&-H_0^6 (32 \k^2 (x-1)^2+63 \k+63)+9 H_0^9 (7+7 \k+4 \k^2 (x-1)^2)+H_0^{12} (45+45 \k\\
&+32 \k^2 (x-1)^2)
-H_0^{15} (45 \k+45-2 \k^2 (x-1)^2))\,.
\end{split}
\eqlabel{c26}
\end{equation}
\begin{equation}
\begin{split}
\calc_{27}=&\frac{\k^2 H_0^8 m^2 (1+\k+H_0^3 (\k^2 (x-1)^2-2-2 \k)+H_0^6 (1+\k))}
{6(1+\k)^2 (H_0^6-1)^2 (H_0^3-1) (2+H_0^3)}\,.
\end{split}
\eqlabel{c27}
\end{equation}
\begin{equation}
\begin{split}
\calc_{31}=&\frac{4 \k (1-x)}{\sqrt{1+\k} (H_0^6-1)}\,.
\end{split}
\eqlabel{c31}
\end{equation}
\begin{equation}
\begin{split}
\calc_{32}=&\frac{6 \b \k (1-2 H_0^3) (1-x)}{\sqrt{1+\k} (H_0^6-1) (H_0^3-1) H_0}\,.
\end{split}
\eqlabel{c32}
\end{equation}
\begin{equation}
\begin{split}
&\calc_{33}=-\frac{1}{(x-1) (H_0^6-1)^2 (1+\k)}(1+\k+4 H_0^3 \k^2 (x-1)^2-2 H_0^6 (\k+1+2 (x-1)^2 \k^2)
\\
&+4 H_0^9 \k^2 (x-1)^2+H_0^{12} (1+\k))\,.
\end{split}
\eqlabel{c33}
\end{equation}
\begin{equation}
\begin{split}
\calc_{34}=&-\frac{8 \k^3 H_0^3 (x-1)^2}{(1+\k)^{3/2} (H_0^6-1)^2 (H_0^3+1)}\,.
\end{split}
\eqlabel{c34}
\end{equation}
\begin{equation}
\begin{split}
\calc_{35}=&\frac{8 \k^3 H_0^2 \b (x-1)^2 (-4 H_0^3+4 H_0^6+2+H_0^9)}{(1+\k)^{3/2} (H_0^3-1) (H_0^6-1)^3}\,.
\end{split}
\eqlabel{c35}
\end{equation}

\section{Sound of $\caln=4$ plasma via the new technique}\label{sbh}
The  technique for computing the holographic sound wave dispersion relation developed in 
section \ref{new} is rather complicated. Thus, we believe that it warrants a simple explicit example. 
This example is being provided by an  $AdS_5$ Schwarzschild black hole, holographically dual to 
strongly coupled $\caln=4$ plasma at finite temperature 
and zero chemical potentials. 
The latter is realized as a special case of the 
background \eqref{back} with 
\begin{equation}
A\equiv 0\,,\qquad \phi\equiv 1\,,\qquad \chi\equiv 0\,.
\end{equation}
In this case the only gauge invariant fluctuations are those of $z_{H,0}$ and $z_{H,1}$. 
The analog of \eqref{order0} is:
\begin{equation}
0=z_{H,0}''+\frac{\b_{12}+3x^2-6x+1}{(1-x)(x^2-2x+3-\b_{12})}\ z_{H,0}'+\frac{4}{x^2-2x+3-\b_{12}}\ z_{H,0}\,,
\eqlabel{order0s}
\end{equation}
and the analog of \eqref{order1} is 
\begin{equation}
\begin{split}
&0=z_{H,1}''+\frac{\b_{12}+3x^2-6x+1}{(1-x)(x^2-2x+3-\b_{12})}\ z_{H,1}'+\frac{4}{x^2-2x+3-\b_{12}}\ z_{H,1}+\calj_H\\
&\calj_H=-\frac{2 \b_{12}^{1/2} }{\sqrt{3}(x-1) (x^2-2 x+3-\b_{12})^2}\ (\b_{12}^2+4 \b_{12} x-2 \b_{12} x^2-6 \b_{12}
-4 x^3+8 \b_2 x\\
&+x^4+9+10 x^2-12 x-4 \b_2-4 \b_2 x^2)\ z_{H,0}' 
+\frac{ 4\b_{12}^{1/2} (x^2-2 x-\b_{12}-2 \b_2+3)}{\sqrt{3}(x^2-2 x+3-\b_{12})^2}\ z_{H,0}\,.
\end{split}
\eqlabel{order1s}
\end{equation}
Note that since we expect $\b_{12}=1$, there are no poles in the connection coefficients in \eqref{order0s}
inside the range of integration $x\in (0,1)$ --- the would-be poles are at 
\begin{equation}
0=x^2-2 x+3-\b_{12}\,.
\eqlabel{poleposs}
\end{equation}
Nonetheless, $\calj_H$ in \eqref{order1s} contains factors of $(x^2-2 x+3-\b_{12})^2$
in the denominators of the connection coefficients. As explained in section \ref{new} 
their origin can be traced back to the hydrodynamic expansion \eqref{sdisp}.  

According to discussion around \eqref{dervalue}, solving \eqref{order1s} while treating $\b_{12}$ 
as a free parameter and imposing the horizon boundary condition as in \eqref{bc} computes $\hz_{H,0}$:
\begin{equation}
\hz_{H,0}=\frac{1-\b_{12}+2x-x^2}{2-\b_{12}}\qquad \Longrightarrow\qquad \qquad 
\calz_0(\b_{12})=\frac{1-\b_{12}}{2-\b_{12}}\,.
\eqlabel{hz0}
\end{equation}
Solving 
\begin{equation}
\calz_0(\b_{12}=\b_{12}^\star)=0\qquad  \Longrightarrow\qquad \b_{12}^\star=1\,,
\eqlabel{sz0}
\end{equation}
determines the speed of sound $c_s^2=\frac{\b_{12}^\star}{3}=\frac 13$, and following \eqref{ident} computes $z_{H,0}$:
\begin{equation}
z_{H,0}=\hz_{H,0}\bigg|_{\calz_0(\b_{12})=0}=2x-x^2\,.
\eqlabel{zh0s}
\end{equation}
Using \eqref{hz0} we compute 
\begin{equation}
\calz_0'=\frac{d}{d\b_{12}}\ \calz_0(\b_{12})\bigg|_{\calz_0(\b_{12})=0}=-1\,.
\eqlabel{z0ps}
\end{equation}
To proceed further, we introduce $\tz_{H,1}$ following \eqref{breakz}. As explained around equation 
\eqref{defjh}, $\tz_{H,1}$ satisfies the following equation
\begin{equation}
\begin{split}
&0=\tz_{H,1}''+\frac{3x^2-6x+2}{(1-x)(x^2-2x+2)}\ \tz_{H,1}'+\frac{4}{x^2-2x+2}\ \tz_{H,1}+\tilde{\calj}_H\\
&\tilde{\calj}_H=\frac{2}{\sqrt{3}(1-x)}\ z_{H,0}' 
+\frac{ 4}{\sqrt{3}(x^2-2 x+2)}\ z_{H,0}\,,
\end{split}
\eqlabel{order1st}
\end{equation}
where we substituted $\b_{12}=\b_{12}^\star=1$.
Notice that $\tilde{\calj}_{H}$ does not contain factors of $(x^2-2x+2)^2$ in the denominators of the 
connection coefficients anymore.
Using \eqref{zh0s} we can solve for $\tz_{H,1}$, subject to the horizon boundary condition \eqref{bc}:
\begin{equation}
\tz_{H,1}=\frac{2}{\sqrt{3}}(2x-x^2)-\frac{2}{\sqrt{3}}\qquad \Longrightarrow\qquad \calz_1=-\frac{2}{\sqrt{3}}
\,.
\eqlabel{z1s}
\end{equation}
Finally, from \eqref{b2def} we compute the attenuation coefficient
\begin{equation}
\b_2=1\,,
\eqlabel{b2s}
\end{equation}  
which is the expected answer leading to a zero bulk viscosity for the $\caln=4$ 
SYM.

\end{document}